\documentclass[aps,prd,notitlepage,reprint,onecolumn,showpacs,superscriptaddress,nofootinbib,floatfix]{revtex4-1}

\usepackage{amsmath,amssymb,graphicx}
\usepackage{mathrsfs}
\usepackage{bbm}
\usepackage{bbold}
\usepackage{slashed}
\usepackage[normalem]{ulem} 
\usepackage{tikz}
\usepackage{verbatim}

\newcommand{\be}{\begin{equation}}
\newcommand{\ee}{\end{equation}}
\newcommand{\bi}{\begin{itemize}}
\newcommand{\ei}{\end{itemize}}
\newcommand{\bea}{\begin{eqnarray}}
\newcommand{\eea}{\end{eqnarray}}

\newcommand{\ud}{\mathrm{d}}
\newcommand{\LCm}{{\scriptscriptstyle -}} 
\newcommand{\LCp}{{\scriptscriptstyle +}}
\newcommand{\LCpm}{{\scriptscriptstyle \pm}}

\newcommand{\LCperp}{{\scriptscriptstyle \perp}}
\newcommand{\sfp}{{\sf p}}
\newcommand{\sfP}{{\sf P}}
\newcommand{\sfq}{{\sf q}}
\newcommand{\sfx}{{\sf x}}

\newcommand{\sfk}{{\sf k}}

\newcommand{\bra}[1]{\langle\,#1\,|}          
\newcommand{\ket}[1]{|\,#1\,\rangle}          

\DeclareMathSymbol{\mlq}{\mathord}{operators}{``}
\DeclareMathSymbol{\mrq}{\mathord}{operators}{`'}

\begin{document}

\title{Mode truncations and scattering in strong fields}

\author{Tom Heinzl}
\email{theinzl@plymouth.ac.uk}
\affiliation{Centre for Mathematical Sciences, University of Plymouth, PL4 8AA, UK}
\author{Anton Ilderton}
\email{anton.ilderton@plymouth.ac.uk}
\affiliation{Centre for Mathematical Sciences, University of Plymouth, PL4 8AA, UK}
\author{Daniel Seipt}
\email{d.seipt@lancaster.ac.uk}
\affiliation{Lancaster University, Physics Department, Bailrigg, Lancaster LA1 4YW, UK}
\affiliation{The Cockcroft Institute, Daresbury Laboratory, Keckwick Ln, Warrington WA4 4AD, UK}

\begin{abstract}
Truncating quantum field theories to a dominant mode offers a non-perturbative approach to their solution. We consider here the interaction of charged scalar matter with a single mode of the electromagnetic field. The implied breaking of explicit Lorentz invariance prompts us to compare instant-form quantisation and front-form, with the latter yielding significant simplifications when light-front zero modes are included. Using these field theory results we reassess the validity of existing first-quantised approaches to depletion effects in strong laser fields, and propose an alternative interpretation based on the dressing approach to QED and its infra-red structure.
\end{abstract}

\maketitle
\section{Introduction}
Exactly solvable interacting field theories are few and far between. They typically arise in two dimensions, e.g.\ in the guise of conformal field theories~\cite{Belavin:1984vu} or perturbations thereof~\cite{Zamolodchikov:1978xm}. The non-perturbative approaches required can be sophisticated, in terms of both physical insight and mathematical techniques. A simpler, but direct and physically accessible approach, is to consider interactions of, say, quantum particles with \textit{fixed} backgrounds rather than fluctuating quantum fields; there are for instance several electromagnetic backgrounds for which the interaction with quantised matter can be treated exactly, with well-known examples being constant fields, Coulomb fields, and plane waves. See~\cite{Bagrov:1990xp} for a review and~\cite{Heinzl:2017zsr} for recent progress. The plane wave case in particular has attracted attention as a model for the strong electromagnetic fields of an intense laser, see~\cite{Seipt:2017ckc} for a recent review.

The Dirac equation can be solved exactly in a background plane wave, yielding first quantised wavefunctions describing the interaction of single particles with a classical plane wave. A background field is of course an idealisation, and preferably one would want all fields to be fully dynamical~\cite{Yang:2017xyh,Ilderton:2017xbj}. In pursuit of this the question has been asked of what happens if, in the Dirac equation, the plane wave background is replaced by an operator-valued plane wave~\cite{Berson,Bergou:1980cp,Filipowicz:1985,Bagrov:1990xp,Skoromnik:2014}; it transpires that the Dirac equation can still be solved, yielding states which are in the literature taken to describe the dynamics of a single particle interacting with a quantised laser field.

Here we will examine such systems, which are truncated in both particle content and photon momentum, from a field theory point of view. We will begin with an interacting field theory, and treat it not with ordinary perturbation theory but by using an alternative split of the Hamiltonian into `free' and `interacting' parts, in which the former contains nontrivial interactions with a single (electromagnetic) mode. The resulting system may be mathematically simple, but its physical interpretation is non-trivial, and we will be forced to reevaluate the connection to intense laser physics. 

This paper is organised as follows. Immediately below we review some literature on single mode interactions in the context of strong field physics. In Sect.~\ref{SECT:YUKAWA} we consider scalar Yukawa theory as a toy model instead of QED, because dropping spin and polarisation greatly clarifies the presentation. We reduce the (scalar) photon to a single momentum mode and quantise. After confronting the issue of light-front zero modes, we obtain an exactly solvable system. In Sect.~\ref{SECT:INTERP} we give a physical interpretation of the system in terms of dressed particles, and with this return to strong field physics in Sect.~\ref{SECT:LAS}. We re-examine the interpretation usually ascribed to the solutions of the `operator-valued' Dirac equation, show how the background field limit is obtained, and how to go beyond it by including effects due to beam depletion. We conclude in Sect.~\ref{SECT:CONCS}.

\subsection{Review} \label{SECT:REVIEW}
The idea that single (bosonic) modes of a quantum field may dominate the physics of a given system goes back at least to Bogoliubov's explanation of superfluidity~\cite{Bogoliubov:1947}, where a single zero-momentum mode leads to a highly populated, macroscopic ground state, explaining the unusual features of superfluids. In the context of asymptotically free theories such as Yang-Mills, one can separate the zero-momentum modes from the rest by working in a small physical volume~\cite{Luscher:1982ma}. This allows for an approximate analytic calculation of the glue ball spectrum~\cite{Luscher:1983gm}, despite the fact that the classical zero-momentum dynamics (`Yang-Mills mechanics'~\cite{Matinyan:1981dj}) is not integrable, but rather chaotic~\cite{Nikolaevskii:1983,Biro:1994bi}.

However, the most common electrodynamical system with a highly dominant mode, i.e.\ a highly occupied state, is the laser. By the correspondence principle, such a state should be well described by a classical field, hence the relativistic interaction of charges with such a macroscopic mode are often treated as background field problems. Such discussions employ the ``Volkov solution''~\cite{Volkov}, which is the exact solution to the Dirac equation in a background plane wave, that is a transverse potential $A^\mu_\text{ext}(k.x)$ with light-like wave vector $k^2=0$:
\be\label{Dirac}
	\big(i \slashed{\partial} - e \slashed{A}_\text{ext}  - m \big) \psi =0 \;.
\ee
The Volkov solution is an essential ingredient in describing intense laser-matter interactions. Consider the calculation of scattering amplitudes in QED with an additional background plane wave modelling the laser. Interactions between particles and photons are treated in perturbation theory as normal, but the coupling $a_0\sim eA_\text{ext}/m$ to the external field is taken to be strong, $a_0>1$. This coupling must therefore be treated exactly, or at least outside of perturbation theory. Using this ``Furry expansion''~\cite{Furry:1951zz,Fradkin:1981sc} of the $S$-matrix, the Volkov solutions appear as external leg (asymptotic particle) wavefunctions through LSZ reduction of the propagator, which is itself the inverse of the Dirac operator in the background~\cite{Ilderton:2012qe,Lavelle:2015jxa,Seipt:2017ckc,DiPiazza:2018ofz}. Thus the presence of a background field leads to modified Feynman rules to be used in strong field physics, with these new rules expressed in terms of the Volkov solutions.

Returning to the Dirac equation, one may be interested in the validity of, and corrections to, the assumption that the electromagnetic field is a background. This is the case in the context of back-reaction, i.e.~when the background field approximation becomes invalid due to, say, depletion effects. This has prompted, in the literature, the question of what happens if the classical plane wave in the Dirac equation is replaced by a quantised, or operator valued, field. This field is for the most part taken to contain a single frequency mode, i.e.~to be a quantised generalisation of a \textit{monochromatic} plane wave.  One replaces in the Dirac equation (\ref{Dirac}),
\be\label{K2K2}
	A^\mu_\text{ext}(k.x) \to  
	{\bar\epsilon}^\mu a^\dagger e^{i k.x} + 
	\epsilon^\mu a e^{-ik.x} \;,
\ee
in which $a$ is the single mode annihilation operator, and $[a,a^\dagger]=1$. It is still possible to solve the Dirac with this extension, as first shown in~\cite{Berson}. The solutions yield states which live in the photon Fock space. These states have been interpreted as a quantum generalisation of the Volkov solution, allowing one to, in principle, treat the laser fully quantum mechanically and so go beyond the background field approximation. However, these generalised Volkov solutions are `first-quantised' with respect to the matter sector of Hilbert space, thus it is not immediately clear what their connection with QFT is, nor how they should be used in scattering calculations. Further, a background plane wave \textit{already has a fully quantised description in terms of coherent states}~\cite{Kibble:1965zza,Frantz,Gavrilov:1990qa,Ilderton:2017xbj}. We will address these issues below, and show that while there are structural similarities between the Volkov solution and the generalised states, the physics of the latter is very different, and substantially richer than described in the literature to date.

\section{A toy model of QED with a single EM mode} \label{SECT:YUKAWA}
We adopt here a toy model of QED in which we drop spin, gauge and polarisation degrees of freedom; this allows us to uncover the important structures and address non-trivial physical questions without the unrevealing spin and polarisation corrections which can make existing investigations of single-mode theories cumbersome. Consider then the Yukawa coupling of a complex scalar $\phi(x)$ to a real scalar $A(x)$, with Lagrangian
\be\label{enkel-L}
	\mathcal{L} = \partial\phi^\dagger. \partial \phi - m^2\phi^\dagger \phi + 
	\frac{1}{2}\partial A. \partial A - e A \phi^\dagger \phi \;.
\ee
The three-point Yukawa vertex mimics that of QED, hence momentum conservation rules are the same. Pursuing the analogy with QED we refer to $\phi(x)$ as the matter field, and its modes as the electron and positron, and to $A(x)$ as describing the photon.  This model is sometimes referred to as the Nelson model~\cite{Nelson:1964}, though this was first introduced with nonrelativistic matter (`nucleon') fields. The theory (\ref{enkel-L}) also serves as a toy model for addressing the infrared problem, see e.g.~\cite{Dybalski:2017mip}, references therein, and below.

We will work in the Hamiltonian formalism. We want to single out the interaction with a single (preferred) photon mode with momentum $k_\mu$. To do so we will split the Hamiltonian into `free' and `interacting' parts by introducing projectors $\mathbb P$ and  $\mathbb Q$ (which sum to unity), in which $\mathbb P$ selects out the single mode and interactions with it. One motivation for this approach is that there may be circumstances in which the eigenstates of the new `free' Hamiltonian are a better starting point for perturbative calculations than the original free particle states~\cite{Dietrich:2006fw}.

\begin{figure}[t!]
\includegraphics[height=0.3\textwidth]{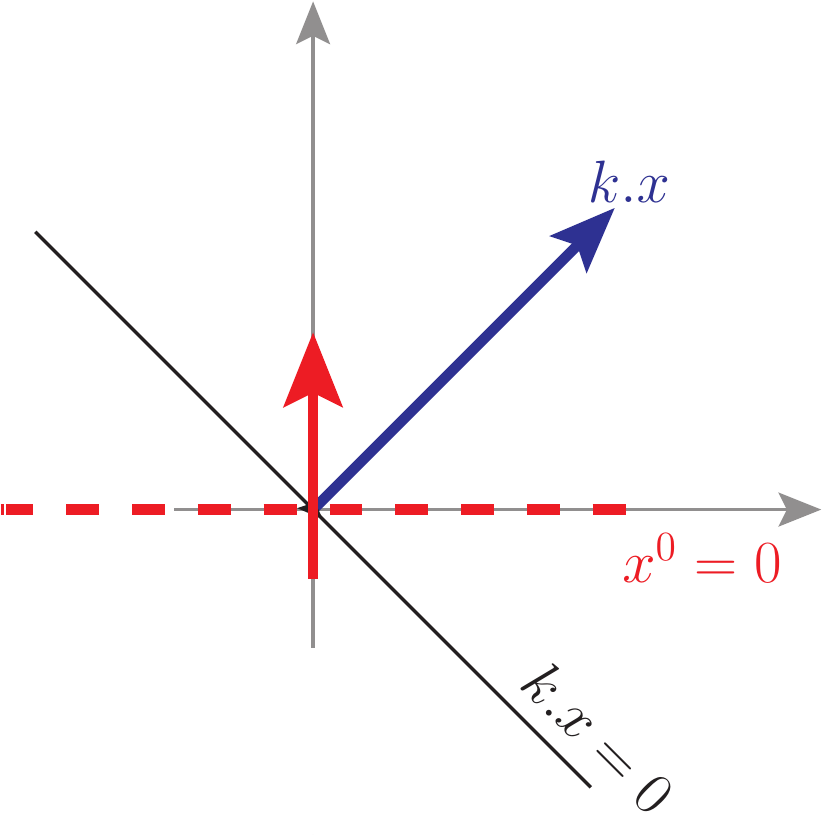}
\raisebox{-6pt}{\includegraphics[height=0.315\textwidth]{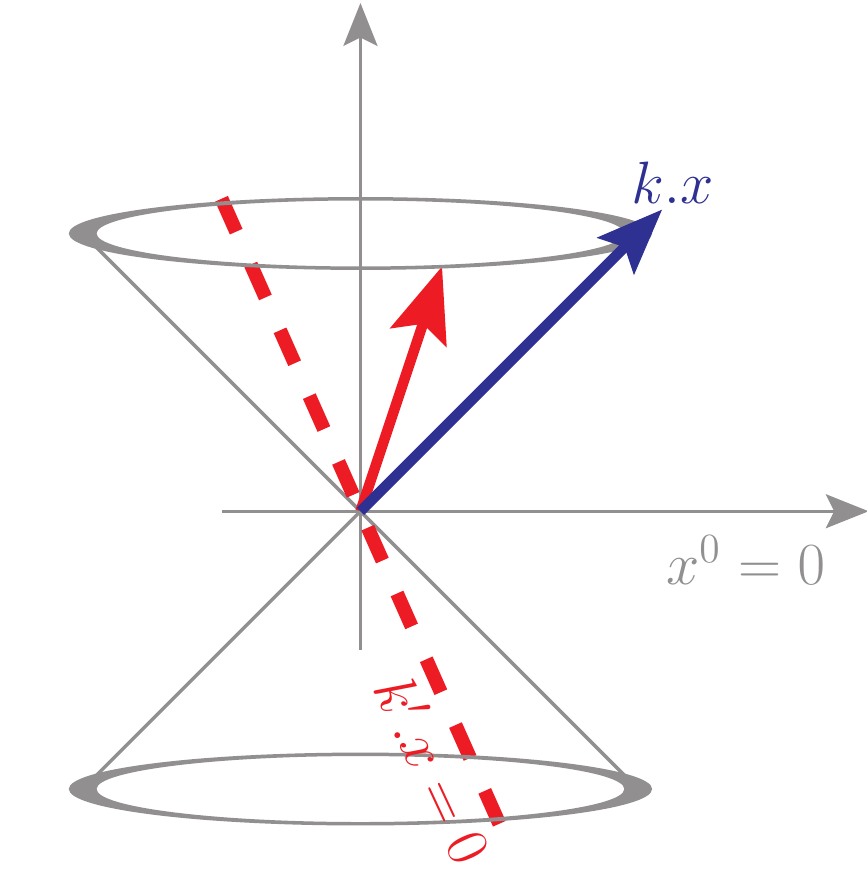}}
\includegraphics[height=0.3\textwidth]{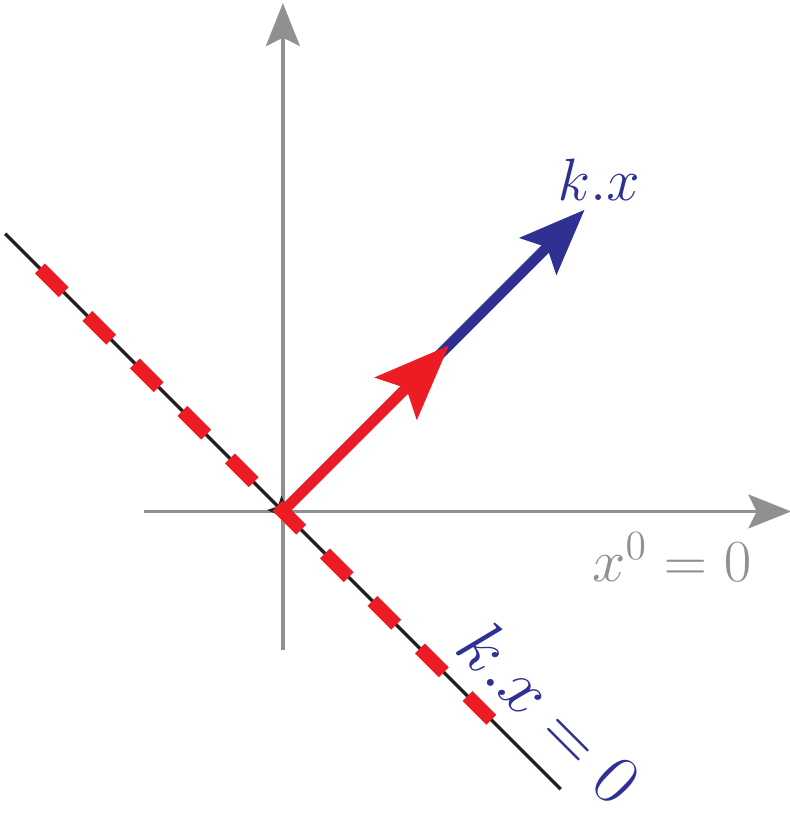}
\caption{\label{FIG:3} The three time directions (red/short) lines and quantisation surfaces (red/dashed lines) used to quantise the theory with a single massless mode of momentum $k$ (blue/long arrow). Left to right: instant form, front form, and front form again but in which the time direction coincides with $k.x$.}
\end{figure}

Clearly our separation will break explicit Poincar\'e invariance. It is then an interesting question of whether or not the choice of time, or quantisation surface, leads to different physics. There are three basic possible choices for the time evolution parameter~\cite{Dirac:1949cp}, each with its own Hamiltonian: the instant form, the front form, and the point form. We will focus on the first two, in which we can still discern three options. The first is the instant form (first panel of Fig.~\ref{FIG:3}), in which $x^0$ is the time direction, and the quantisation surface is $x^0=0$. Since the single mode momentum $k_\mu$ is a light-like vector, $k.x$ lies on the light-cone and hence in this setup the theory has two preferred directions, parametrised by the coordinates $x^0$ and $k.x$. Proceeding this way, and as detailed in the appendix, we find no closed form solution to the resulting Schr\"odinger equation, nor any connection to (the scalar Yukawa analogue of the) first quantised literature results.

In the second setup (second panel of Fig.~\ref{FIG:3}) we use the front form. We choose a light-like time direction $k'.x$ with $k'.k'=0$ and $k'_\mu \nparallel k_\mu$, so $k.k'\not=0$. There are again two preferred directions, but (see the appendix) one can construct some eigenstates of the resulting system exactly. Once again, though, they do not seem to have anything in common with literature results. Our third and final setup (third panel of Fig.~\ref{FIG:3})  again uses the front form, but where the time direction is coincident with $k.x$. Since there is now only one preferred direction in play, we might expect a simplification, and this will indeed be the case: the single mode theory quantised in this way is exactly solvable, and it is precisely this setup which connects to existing results. We turn to this now.

\subsection{Front form quantisation with zero modes}\label{SECT:ZM}
We wish to quantise our theory in the front form, as reviewed in~e.g.~\cite{Burkardt:1995ct,Brodsky:1997de,Heinzl:2000ht}, such that the time direction and single-mode momentum direction are coincident, see Fig.~\ref{FIG:3}, right panel.  We define coordinates $x^\LCpm = x^0 \pm x^3$, $x^\LCperp=(x^1,x^2)$, and take $x^\LCp$ as the time. Momenta are $p_\LCpm  =(p_0 \pm p_3)/2$ and $p_\LCperp=\{p_1,p_2\}$. With this, the single photon mode of interest carries zero light-front momentum, $k_\LCm=k_\LCperp=0$, but has non-zero light-front energy, $k_\LCp=\omega$.

We immediately face the problem that quantising such degrees of freedom in the front form is challenging~\cite{Heinzl:2003jy,Collins:2018aqt}.  One difficulty is that the on-shell momentum measure in light-front coordinates $\sim \ud^4 k\delta( k^2 ) \theta( k_\LCp) = \ud^2 k_\perp \ud k_\LCm/2 k_\LCm$ is not complete, and as a result not all modes are included in the Fock space expansion of the field operators. The missing modes are exactly the light-front zero modes with $k_\LCm=0$ which we wish to retain. We therefore have to find a proper mode expansion that includes the photon zero modes, which are usually neglected.
A second difficulty is that there is no free Hamiltonian when quantising zero modes on the initial surface $x^\LCp=0$, within which zero modes propagate \emph{instantaneously} \cite{Heinzl:2003jy}. In order to define the light-front Hamiltonian, i.e.~the generator of the light-front time evolution, we adopt a method used to quantise the Schwinger Model~\cite{Schwinger:1962tn} on the light-front~\cite{McCartor:1988bc}, see also~\cite{diss:Hornbostel,Bogoliubov:1990}.

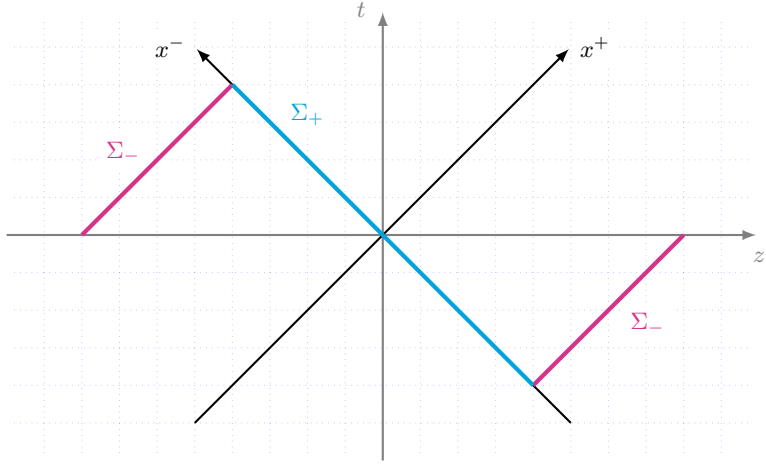
\begin{figure}[!t]
\centering\begin{tikzpicture}[node distance=25mm,
								gridlines/.style={,very thin,dotted,blue!40},
		    					axis/.style={->, thick, shorten >=1pt,>=latex,gray},                      
		    					lfaxis/.style={->, thick, shorten >=1pt,>=latex}                      
                     ]
\draw[gridlines,step=0.5] (-4.9,-2.9) grid (4.9,2.9);
 \draw[axis] (-5,0) -- (5,0) node [below=3pt,fill=white] {$z$}  ;
\draw[axis] (0,-3) -- (0,3) node [left=3pt,fill=white]     {$t$}  ;
\draw[lfaxis] (-2.5,-2.5) -- (2.5,2.5) node [right,fill=white] {$x^+$};
\draw[lfaxis] (2.5,-2.5) -- (-2.5,2.5) node [left,fill=white] {$x^-$};
\draw (-3,1) node [magenta!90!black,above=3pt,left=3pt,fill=white] {$\Sigma_-$};
\draw (3,-1) node [magenta!90!black,below=5pt,right=5pt,fill=white] {$\Sigma_-$};
\draw (-1,1) node [cyan!90!black,above=10pt,fill=white] {$\Sigma_+$};
\draw[magenta!90!black, ultra thick] (-4,0)  -- (-2,2); 
\draw[magenta!90!black, ultra thick]  (2,-2)  -- (4,0);
\draw[cyan!90!black, ultra thick]  (-2,2) -- (2,-2);
\end{tikzpicture}
\caption{\label{FIG:McCartor} Quantisation surfaces, $\Sigma_\LCpm$, used to define a light-front Hamiltonian containing zero modes. $\Sigma_\LCp: x^\LCp = 0$, $\Sigma_\LCm: x^\LCm = \pm L$ (piecewise constant).}
\end{figure}

From here we use sans-serif font for the light-front position and momentum three-vectors $\sfx=(x^\LCm,x^\LCperp)$ and $\sfp=(p_\LCm,p_\LCperp)$. We also define an integral measure and delta function by $\ud\sfp := \ud^2p_\LCperp \ud p_\LCm \theta(p_\LCm)$ and $\delta(\sfp - \sfq)  := \delta(p_\LCm-~q_\LCm)\delta^2(p_\LCperp-q_\LCperp)$.  Imagine the system to be quantised in a box of length $L$. The canonical momenta $P^\mu$ are defined by integrating the stress-energy-momentum tensor $\Theta^{\mu\nu}$,
\be
	\Theta^{\mu\nu} =\partial^\mu\phi^\dagger \partial^\nu\phi  + \partial^\nu\phi^\dagger \partial^\mu\phi + \partial^\mu A \partial^\nu A  - g^{\mu\nu} \, \mathcal L  \;,
\ee
over McCartor's surface shown in Fig.~\ref{FIG:McCartor}~\cite{McCartor:1988bc}, which yields the two contributions
\begin{align}
P_\mu = \int_{\Sigma_+} \! \ud x^- \ud^2 x^\perp\, \Theta_{\LCm\mu} 
			+\int_{\Sigma_-} \! \ud x^+ \ud^2 x^\perp\, \Theta_{\LCp\mu} \;.
\end{align}
In particular, the light-front Hamiltonian, $P_\LCp$, becomes
\begin{align}\label{bracing}
  P_\LCp  &= P_\LCp^\phi 
		+  \underbrace{ \frac{1}{4} \int \! \ud x^- \ud^2 x^\perp \: 
		(\nabla_\perp A)^2 }_{\displaystyle \equiv   
		P_\LCp^{A,\mathrm{nzm}} }
		+  \underbrace{ \frac{e}{2} \int \! \ud x^- \ud^2 x^\perp \: 
		A \phi^\dagger\phi     }_{\displaystyle \equiv P_\LCp^{\mathrm{int}}}
       +   \underbrace{  \int \! \ud x^+ \ud^2 x^\perp \: 
       (\partial_\LCp A)^2 }_{\displaystyle \equiv P_\LCp^{A,\mathrm{zm}}  }
\end{align}
The first three terms are, respectively, the free Hamiltonian for $\phi$,
the free Hamiltonian for the non-zero modes of $A$, and the interaction Hamiltonian. The fourth term derives from the integral over $\Sigma_\LCm$ and gives the free zero mode Hamiltonian of $A$. This is absent for a single quantisation hyper-surface, $x^\LCp=0$. To proceed, we consider the more familiar terms first.
%
%
The (normal ordered) free matter Hamiltonian, $P^\phi_\LCp$, has contributions from both $\Sigma_\LCp$ and $\Sigma_\LCm$. We ensure the absence of matter field zero modes by choosing Dirichlet boundary conditions. Consequently, $P^\phi_\LCp$ reduces to the standard light-front expression in the infinite-volume limit,  
\begin{align}
 	P^\phi_\LCp &= 
 			\int \! \ud\sfp \, p_\LCp 
 							 ( b_\mathsf{p}^\dagger b_\mathsf{p} + d_\mathsf{p}^\dagger d_\mathsf{p}) \,, \quad \text{with}�\quad p_\LCp = \frac{p_\perp^2 + m^2}{4 p_\LCm} \;, \quad{\text{and}}\quad [b_\sfp,b^\dagger_\sfq]=[d_\sfp,d^\dagger_\sfq]=\delta(\sfp-\sfq) \;.
\end{align}
In other words, the Fock expansion of $\phi(x)$ is as normal (see~\ref{FOCK-EXP-LF} in the appendix). On the other hand, the mode expansion for the photon field $A$ should contain both non-zero and zero modes, which we make explicit by expanding
\begin{eqnarray}
  A(x) &=& \int \! \frac{\ud \sfk}{\sqrt{(2\pi)^32k_\LCm}} \: (a_\sfk e^{-ik.x} + 
  a^\dagger_\sfk e^{ik.x} )
  +  \frac{1}{|L_\LCperp|} \int\limits_0^\infty\! \frac{\ud k_+}{\sqrt{(2\pi)2k_\LCp}}\, 
  (\alpha^\dagger_{k} e^{ik_+x^+} + \alpha_{k} e^{-ik_+x^+}) \nonumber \\ 
  &=:& A_\mathrm{nzm}(x) +  A_\mathrm{zm}(x^+) \;. \label{A.ZM.NZM}
\end{eqnarray}
The normalisation of Fock operators in the zero-mode term, $A_\mathrm{zm}$, will become clear in a moment. The free Hamiltonian for the \textit{non-zero} modes of $A$ receives contributions only from $\Sigma_\LCp$ which returns the usual result
\be
	P^{A,\rm nzm}_\LCp = \int \! \ud\sfk \, \frac{k_\perp^2}{4 k_\LCm}  \:
 							 a^\dagger_\sfk a_\sfk \;, \quad \text{with} \quad [a_\sfp,a^\dagger_\sfq]= \delta(\sfp-\sfq) \;.
\ee
Turning to the zero-mode term, $A_\mathrm{zm}$, we note that the associated Hamiltonian,
\begin{align}
	P_\LCp^{A,\mathrm{zm}} = 
	\int \! \ud x^+ \ud^2 x^\perp (\partial_\LCp A_\mathrm{zm})^2 
	&=
	\int\limits_0^\infty\! \ud k_\LCp k_\LCp\, \alpha^\dagger_k \alpha_k \; , 
\end{align}
describes a continuum of zero modes with a free Hamiltonian in the form expected for modes of energy $k_\LCp$, hence we impose the commutators
\be\label{alpha-comm}
	[\alpha_{k_+} , \,  \alpha^\dagger_{p_+} ] = \delta(k_\LCp - p_\LCp)  \quad \mbox{and} 
	\quad [a_\sfk, \alpha^\dagger_{p_+}] = 0 \; ,
\ee
(explaining the normalisation chosen in (\ref{A.ZM.NZM})), which is equivalent to imposing the equal-$x^\LCm$ field commutator
\be
  \big[ A_\text{zm}(x^\LCp) , \; 2 \partial_\LCp A_\text{zm}(y^\LCp) \big] = \frac{1}{L_\LCperp^2}
  \, \delta(x^\LCp - y^\LCp) \; .
\ee
The interaction Hamiltonian in (\ref{bracing}) can be split into two parts, corresponding to the interaction of $\phi$ with zero modes and non-zero modes of $A$, respectively. The zero-mode part is	
\be \label{eq:light-front-interaction-zero-mode}
	P_\LCp^\mathrm{int,zm} = \frac{e}{2} A_\mathrm{zm}(x^\LCp=0) \int \! \ud x^- \ud^2 x^\LCperp \,  
	\phi^\dagger \phi 
	= e A_\mathrm{zm}(x^+=0) \int \!  \frac{ \ud \sfp }{4p_-} \:
	( b_\mathsf{p}^\dagger b_\mathsf{p} + d_\mathsf{p}^\dagger d_\mathsf{p}) \;,
\ee
where $A$ is taken at $x^\LCp=0$ because we integrate over the surface $\Sigma_\LCp$, see Fig.~\ref{FIG:McCartor}.

Let us collect all terms containing the free dynamics of the matter field $\phi$, the photon zero mode(s), and their mutual interactions (as the electron and positron are automatically decoupled, we drop the latter):
\be \label{eq:Pminus:0}
		 P^\phi_\LCp + P_\LCp^{A,\rm zm} + P_\LCp^\mathrm{int,zm} = 
		 	\int \! \ud\sfp \, p_+  \:  b_\mathsf{p}^\dagger b_\mathsf{p}  + \int\limits_0^\infty\!\ud k_\LCp \, k_\LCp \: \alpha_{k_\LCp}^\dagger \alpha_{k_\LCp} 
		+ \frac{e}{2|L_\LCperp|}\int\limits_0^{\infty}\! \ud k_+\frac{\alpha_k + \alpha^\dagger_k }{\sqrt{(2\pi) 2k_\LCp}}  \int \!  \frac{ \ud \sfp }{4p_-} \:  b_\mathsf{p}^\dagger b_\mathsf{p}  	\,.		 
\ee
All terms including non-zero modes of the photon are to be treated as part of the interaction Hamiltonian, which will be reintroduced later. To make this concrete we explicitly reduce the photon to a single zero mode with $k_\LCp=\omega\not=0$ (but $k_\LCm=k_\LCperp=0$) by inserting into all Fourier transforms of $A(x)$ the resolution of unity $1 = \mathbb{P} + \mathbb{Q}$, where~\cite{Heinzl:1991vd}
\be\label{P-Q-11}
	\mathbb{P} = \frac{\delta(k_\LCp-\omega)}{\delta(0)} \;, \qquad \mathbb{Q}= 1- \mathbb{P} \;.
\ee
Doing so, and rescaling the single mode operator as $\alpha_{\omega} = a \sqrt{\delta(\sf{0})}$ such that $[a,a^\dagger]=~1$, we obtain our `free' Hamiltonian
	\be\label{H-enkelt2-LF}
	H_\text{zm} := \omega a^\dagger a +\int\!\ud{\sf p}\; 
	b^\dagger_\sfp b_\sfp \left[ p_\LCp  +  \frac{g}{4p_\LCm} \big(a + a^\dagger	\big) \right] \;,
	\ee
	%
with the effective coupling $g = e / \sqrt{2 \omega L^+ L_\perp^2 }$. This is, finally, the zero-mode Hamiltonian we wish to study.

There are two immediate consequences of, and significant simplifications due to,  singling out a zero mode.  First, the Hamiltonian contains no momentum changing terms. Second, all (matter particle) number-changing terms drop out of the Hamiltonian due to longitudinal momentum conservation~\cite{Maskawa:1975ky,Heinzl:2000ht}, hence the different $n$-particle sectors of Fock space are decoupled \textit{automatically}. (In both the instant form of quantisation, and in the `non-coincident' front form, one must remove at least some such terms by hand, see the appendix.) Hence the Hamiltonian becomes diagonal in both (matter) particle number and momentum.  This theory is simple, and exactly solvable, as we show below. However, the question of what physics it describes is non-trivial. The remainder of this paper is given over to addressing this question and connections with different theories and literature approaches.

\section{Physical interpretation}\label{SECT:INTERP}

\subsection{Bare vs. physical dressed states} \label{SECT:DRESSING}
Recalling the definition of the displacement operator, $D(z) = \exp(a^\dagger z - a z^\dagger)$ for $z\in\mathbb{C}$, we observe that the Hamiltonian (\ref{H-enkelt2-LF}) can be diagonalised using an \textit{operator valued} displacement operator $D(\sigma)$, where the argument $\sigma$ is the self-adjoint operator
\be
	\sigma := \int\!\ud \sfp\ g_p  b^\dagger_\sfp b_\sfp  \;, \qquad g_p := \frac{g}{4 \omega p_\LCm} = \frac{g}{2k.p} \;.
\ee
$D(\sigma)$ acts on $a$-modes as a translation operator, but on $b$-modes as a scaling operator: 
\be\begin{split}
	D(-\sigma) a D(\sigma) = a + \sigma  &=: A  \;,  \qquad D(-\sigma) b_\sfp D(\sigma) = e^{g_p(a^\dagger-a)}b_\sfp  =: B_\sfp  \;,
\end{split}
\ee
which also defines the mutually commuting operators $A$ and $B_\sfp$. In terms of these $H_\text{zm}$ and $\sigma$ take on the simple~forms
\be\label{H-ZM-FYS}
	H_\text{zm} = \omega A^\dagger A + \int\! \ud \sfp\ p_\LCp B^\dagger_\sfp B_\sfp  - \omega \sigma^2 \;, \qquad \sigma = \int\!\ud \sfp\ g_p  B^\dagger_\sfp B_\sfp \;.
\ee
(A canonical transformation of this form was first used in polaron theory~\cite{Lang:1963} and is routinely used in optomechanics~\cite{Mancini,Bose:1997,Aspelmeyer}.) This suggests that $A$ is a free field while $B_\sfp$ has a peculiar momentum-preserving four-point interaction which, being diagonal in any electron sector, contributes only to energy eigenvalues (see below). The eigenstates of $H_\text{zm}$ may hence be written as standard products of $A^\dagger$ and $B_\sfp^\dagger$ acting on the shifted vacuum $D(-\sigma)\ket{0}$. In fact, the vacuum is \textit{invariant} under the shift, i.e.~$D(-\sigma)\ket{0} = \ket{0}$. Using this, it is easily checked that the eigenstates of $H_\text{zm}$ may equivalently be obtained simply by acting with $D(-\sigma)$ on \emph{free} states. A basis of such eigenstates is
\be \begin{split}
\label{phys-state}
	\ket{\{p_i\};n} := B^\dagger_{\sfp_1}\ldots B^\dagger_{\sfp_N} \frac{(A^\dagger)^n}
	{\sqrt{n!}} \ket{0}  = D(-\sigma) b^\dagger_{\sfp_1} \ldots b^\dagger_{\sfp_N} \ket{n} \;, \quad \text{for} \quad N,n\geq 0 \;. 
\end{split}
\ee
These states have (light-front) energy eigenvalues $E_{\{p_i\},n}$ given by
\be\label{gen-psi-los}
\begin{split}
	E_{\{p_i\},n} = \alpha + n \omega - \beta^2 \omega \;, \quad \text{where} \quad \alpha :=  \sum\limits_{i=1}^N p_{i\LCp} \;, \quad \beta := \sum\limits_{i=1}^N g_{p_i} \;.
\end{split}
\ee
(To be explicit, we can evaluate the action of $D(-\sigma)$ on the $b$-modes and  write $\ket{\{p_i\};n}= b^\dagger_{\sfp_1}\ldots b^\dagger_{\sfp_N} D(-\beta)  \ket{n}$.)

In terms of the $a$ and $b$ modes, we see that free photon states are eigenstates, but free electron states are not; if we place electrons into the system they cause a `back-reaction' which forces the photonic part of state to become partially coherent. The coherence depends on the electron momenta, yet the states are also labelled by a `preferred' photon number $n$, the meaning of which we will explain below. (In quantum optics, such photonic states are called semi-coherent~\cite{Boiteux:1973} or displaced Fock states~\cite{Wuensche:1991}.) But what is the physical interpretation of these eigenstates?

Because the shifted vacuum is the Fock vacuum, hence the interaction in $H_\text{zm}$ does not create particles, and because $\sigma^2$ is diagonal, we see that $B^\dagger_\sfp$ and $A^\dagger$ are the operators which diagonalise the Hamiltonian and we interpret them as creating \textit{physical} electrons and photons, while $b^\dagger_\sfp$ and $a^\dagger$ create particles variably referred to as \textit{bare}~\cite{Henley:1962} or virtual~\cite{Compagno:1995}. For example, using that $[A,B_\sfp]=0$, the number of \textit{physical} photons in the one-electron state $B^\dagger_\sfp\ket{0}$ is, consistent with our interpretation, equal to zero:
\be\label{N-REAL}
	A^\dagger A\: B^\dagger_\sfp\ket{0} = 0 \;.
\ee
However, the expectation value of the number of bare photons in this state is
\be\label{N-VIRT}
		\frac{1}{V}\bra{0}B_\sfp\: a^\dagger a\: B^\dagger_\sfp\ket{0} = g_p^2 \;,
\ee
where the volume factor comes from the delta-function normalisation of the $B$-modes. 
Thus, in terms of the original, bare variables, we interpret the coherent state factor~$D(-\sigma)$ as creating a cloud of virtual photons around the electron; if we could solve the corresponding eigenvalue problem in QED (for all photon modes, not just the light-front zero modes) then a photonic ``dressing'' of electrons would indeed emerge. Part of this dressing would describe, and is responsible for, the Coulomb field of the electron~\cite{Dirac:1955uv,Bagan:1999jf,Bagan:1999jk,Ilderton:2007qy}.

The light-front energy of the eigenstates is, from (\ref{gen-psi-los}), the total energy of the electrons and $n$ photons, reduced by the factor $-\omega\beta^2$. We would expect though that physical electrons and photons in the theory should have the usual on-shell energies~\cite{Lee:1981mf}. It is the quartic self-interaction term, $\sigma^2$, in the Hamiltonian which is responsible for the energy reduction. That $\sigma^2$ appears when we transform to `dressed' operators suggests an interpretation of $\sigma^2$ as generating a kind of binding energy \cite{Henley:1962}, which should be unobservable. This is corroborated by (\ref{N-VIRT}), which shows that $\omega \sigma^2$ gives the energy of the bare photon cloud around the electron. We will argue below that the $\sigma^2$ term should thus be renormalised away; first though we will see what else this term can influence.

\subsection{The S-matrix and dressed states}
\label{SECT:spridning}
We have seen that it is possible to treat part of the scalar Yukawa interaction exactly, namely the coupling of matter to a single mode.  Here we will calculate amplitudes for the scattering of the physical (dressed) electron-photon eigenstates above.  In order to have scattering we reintroduce the so-far neglected modes of $A(x)$ in perturbation theory. The corresponding interaction term is, in terms of the original Hamtilonian $H$, 
\be\label{V-DEF}
	V := -\frac{e}{2} \int\!\ud\sfx \ \phi^\dagger({\sfx}) \phi({\sfx}) A_\mathbb{Q}({\sfx}) \;,
\ee
in which the subscript on $A$ reminds us to include the $\mathbb{Q}$-projection onto the photon momentum modes absent from~$H_\text{zm}$. 

Consider the scattering of an initial state $\ket{i} = b^\dagger_\sfp D(-g_p)\ket{n}$, containing $n$ photons and an electron of momentum $p$, to a final state containing $n'$ photons and an electron of momentum $p'$, together with the emission of an additional photon of momentum $k'_\mu\not= k_\mu$, i.e.~emission into some other mode of the photon field. The final state $\ket{f}$ should then also contain the additional photon mode, so $\ket{f} = b^\dagger_{\sfp'} D(-g_{p'})\ket{n'} \otimes \ket{k'}$. Practically, we would also include the free Hamiltonian for these modes in $H_\text{zm}$. (This would not affect any previous calculation, but would allow us to treat the free evolution of the non-zero modes exactly, as normal.) The $S$-matrix element is
\be\label{SFI-PERT}
	S_{fi} = {\bra{f}} \mathcal{T}_\LCp \exp \bigg[- i \int\!\ud \tau \, e^{i H_\text{zm} x^\LCp} {V} e^{-i H_\text{zm} x^\LCp}\bigg] \ket{i} \;,
\ee
Expanding to lowest order in $V$ we find a simple amplitude supported on the momentum conservation law
\be\begin{split}\label{59}
	p'_\mu + k'_\mu + n' k_\mu -(g^2_{p'}-g_p^2)k_\mu &= p_\mu + n k_\mu \;.
\end{split}
\ee
This is not the expected conservation law for the scattering of physical on-shell particles because of the terms $\sim g^2 k_\mu$ (which come from $\sigma^2$). In particular, take $n'=n=0$, then our process describes the emission of a photon from an electron, but this process should be forbidden by momentum conservation.

We are in the unusual position of having an exactly solvable system; but this opens up the question of how to interpret `nonperturbative' results. The question to address here is what to do with the $\sigma^2$ term. As $\sigma$ describes the coupling of the zero mode of the operator product $\phi^\dagger\phi$ to the photon zero mode, $\sigma^2$ thus describes the interaction of two electrons by exchange of (virtual) zero-mode photons.

Now, in terms of the physical operators, see (\ref{H-ZM-FYS}), each electron number sector and each momentum sector is disjoint; the only effect of the $\sigma^2$ term is then to introduce a ``sector-dependent'' energy shift.  We have seen that this affects scattering processes, through the conservation of lightfront energy~(\ref{59}). There can be no other effect, since the interaction preserves all other quantum numbers (lightfront momentum, particle number, photon number). Hence, in terms of scattering within the zero-mode theory, the action of $\sigma^2$ is \textit{degenerate} with free propagation, or no interaction at all, except for the energy shift. 

Now, the zero-mode Hamiltonian is analogous to that used in solid state systems involving electron-phonon interactions. The corresponding term in the semiconductor system of~\cite{Lang:1963} (describing the coupling of two electrons at different lattice sites, via the exchange of virtual phonons) is omitted from the Hamiltonian because it is smaller than already neglected terms. On the other hand, in ~\cite{Bose:1997} the phase generated by the corresponding term becomes important in the construction of superpositions of  cavity modes.

Thus, how we deal with the $\sigma^2$ term depends on the theory at hand, as discussed in Appendix~\ref{SECT:OPTO}. Here we have a relativistic QFT, so we expect to have to renormalise: even though there are no UV divergences in our theory because of the restrictions on modes and momenta, a finite renormalisation may still be required.

We note that the energy shift caused by $\sigma^2$ is equal to the negative of the energy in the cloud of bare photons around the electrons (\ref{N-VIRT}), which is the ground state (`vacuum') energy in the given electron sector. Given this, and the preceeding discussions, we proceed as follows. We perform a ``sector-dependent renormalisation'', subtracting the ground state energy in each sector. This corresponds simply to removing the $\sigma^2$ term from the Hamiltonian, or equivalently adding the ``missing'' energy of the photonic dressing. As such we are essentially imposing a renormalisation condition of isospectrality, such that particle energies in the interacting theory have the expected (free) spectrum. Doing so, the Hamiltonian becomes free in terms of the physical particles, but the fact that these are dressed still has physical consequences when other modes are reintroduced: we see this by returning to scattering. Evaluating the $S$-matrix element (\ref{SFI-PERT}) explicitly with our prescription gives, to lowest order in the coupling to non-zero modes, but exact in the coupling to zero modes,
\be\label{sfi-sista}
\begin{split}
	S_{fi} &\to 
	 ie\,(2\pi)^4 \delta^4( p' +k' +n' k- p-n k) \bra{n'}D(g_{p'} - g_p)\ket{n} \;.
\end{split}
\ee
The momentum conservation law is now the expected one and the nontrivial overlap is an associated Laguerre polynomial, as first noted by Feynman~\cite{Feynman:1950ir}, and explicitly calculated in~\cite{Cahill:1969it,BB,Schwinger:1977ba,Wuensche:1991}. To be explicit, and to understand the physical meaning of the exponential factor, consider $n=1$ and $n'=0$, i.e.~the scattering of an electron and a photon, in which the photon is absorbed, and another is emitted into a different mode. In other words, Compton scattering. The $S$-matrix element (\ref{sfi-sista}) then becomes
\be\label{sfi-compton}
\begin{split}
	S_{fi} =  ie\,(2\pi)^4 \delta^4( p' +k' - p-k) \bigg(\frac{g}{2k.p} - \frac{g}{2k.p'}\bigg) 
	\exp \bigg[-\bigg(\frac{g}{2k.p} - \frac{g}{2k.p'}\bigg)^2\bigg] \;. 
\end{split}
\ee
Everything preceding the exponential is the lowest order (scalar Yukawa) Compton scattering amplitude as can be derived from the Feynman rules, with a coupling $e$ to non-zero modes and a coupling $g$ to zero modes. The $g$-dependent term is proportional to the classical amplitude for emitting (scalar) photons by an electron changing its momentum from $p$ to $p'$. Of particular interest is the final exponential, which comes from the overlap in (\ref{sfi-sista}) . This looks like an all-orders infra-red correction to Compton scattering, even though the single mode frequency $\omega=k_\LCp$ is not necessarily soft. This is the type of infra-red factor which has been found by Chung~\cite{Chung:1965zza}, Kibble~\cite{Kibble:1968a} and Kulish and Faddeev~\cite{Kulish:1970ut} upon using generalised coherent states such as (\ref{gen-psi-los}) to eliminate infra-red divergences. Such factors also arise in the dressing approach to QED~\cite{Dirac:1955uv}, in which the use of physical charges (fermions dressed by clouds of photons) leads to infra-red finite Green's functions at all orders of perturbation theory, see~\cite{Horan:1998im,Bagan:1999jf,Bagan:1999jk}.

Due to our identification of the physical electron modes as dressed states, the similarity with the dressing approach to QED, and the physically sensible structure of e.g.~the Compton scattering amplitude (\ref{sfi-compton}), we adopt the sector-dependent renormalisation described above from here on. To further corroborate our dressing interpretation, we consider the overlap between bare and dressed states.

\subsection{Lippmann-Schwinger equation}
Non-perturbative equations in QFT rarely have exact solutions. This is true e.g.\ for  Schwinger-Dyson equations (but see~\cite{Bender:2010hf}) and Lippmann-Schwinger equations~\cite{Lippmann:1950zz}, which relate free states to scattering states as reviewed in~\cite{Weinberg:1995mt}. In our model, though, we can solve the Lippmann-Schwinger equation exactly. 

Let $\ket{q}$ be an eigenstate of $H_\text{zm}$, and $\ket{q}_0$ the eigenstate of the free theory with the same energy. (The states will share the same momentum, as this commutes with the interaction.) Then the Lippmann-Schwinger equation is
\be\label{LS-IN-1}
	\ket{q} = \ket{q}_0 + \frac{1}{q_\LCp - H_\text{free} + i \epsilon} (H_\text{zm} - H_\text{free}) \ket{q} \;.
\ee
As we have the eigenstates of $H_\text{zm}$ we can `work backwards' and identify the free field state $\ket{q}_0$ corresponding to a dressed state $\ket{q}$, which we choose to be a single electron eigenstate as defined in (\ref{gen-psi-los}), so $\ket{q} =~D(-g_p)b^\dagger_\sfp\ket{n}$. To identify the corresponding free theory state we first exponentiate the energy denominator in (\ref{LS-IN-1}), and let the resulting exponential of $H_\text{free}$ act on $H_\text{zm}-H_\text{free}$ and the eigenstate $\ket{q}$. This yields the integral representation
\be\label{LS-UT-1}
\begin{split}
	\ket{q}_0 &= \ket{q}+ \int\limits_0^\infty \!\ud s\: e^{-\epsilon s} \frac{\ud}{\ud s} D\big(-g_p e^{-i \omega s}\big) b^\dagger_\sfp \ket{n} \;.
\end{split}
\ee
Integrating by parts, the boundary term kills the `$\ket{q}$' on the right hand side. Inserting a complete set of free states into the remaining integral yields
\be\label{LS-UT-2}
	\ket{q}_0 = \sum\limits_{r=0}^\infty \frac{\epsilon}{\epsilon + i (r-n)} b^\dagger_\sfp\ket{r}\bra{r}D(-g_p) \ket{n} \;.
\ee
The $\epsilon$-dependent term reduces to the representation of $\delta(r-n)/\delta(0)$, which is the Kronecker delta. Hence only the term $r=n$ is selected from the sum. Defining a $Z$-factor via $\sqrt{Z_n}= \bra{n}D(-g_p) \ket{n}$  we find
\be\begin{split}\label{LS-UT3}
	\ket{q}_0 = b^\dagger_\sfp \ket{n}\sqrt{Z_n}\;.
\end{split}
\ee
This is the state we expect if $D(-\sigma)$ is a dressing, i.e.~a cloud of bare, rather than physical, photons around the electrons\footnote{Had we included the $\sigma^2$ term, the spectra of $H_\text{zm}$ and $H_0$ would differ, the former being able to take negative values. This contradicts the assumptions of Lippmann-Schwinger, and in pursuing the calculation we would have found $r+g_p^2-n$ in the denominator of (\ref{LS-UT-2}), i.e.~we would only have been able to connect the interacting state to a free state for $n-g_p^2>0$.}. We note that $Z_n$ is the probability of finding a bare electron plus $n$ bare photons in a physical electron plus $n$ photon state, so that~$Z_n$ indeed looks like a wavefunction renormalisation.
%
%

\section{Scattering in strong fields}\label{SECT:LAS}
Having established the physical content of the zero-mode Hamiltonian, we turn to the connections with previous investigations of scattering in strong fields and the (first-quantised) solutions of the Dirac equation with operator-valued electromagnetic fields, as reviewed in Sect.~\ref{SECT:REVIEW}.

\subsection{First vs. second quantisation}
From the Lagrangian (\ref{enkel-L}), our analogue of the Dirac equation in an external field $A_\text{ext}(x)$ is the Klein-Gordon equation
\be\label{KGV}
	(\partial^2 + m^2) \varphi(x) + g A_{ext}(x) \, \varphi(x) =0 \;.
\ee
As with the Dirac equation we interpret (\ref{KGV}) as defining a single-particle wave function $\varphi(x)$. A standard example of this kind is the Dirac equation in an external Coulomb field, which leads to the eigenstates and energies for the relativistic hydrogen atom, see e.g.~\cite{Bjorken:1964}. Here we are interested in plane wave backgrounds, $A_\text{ext}= A_\text{ext}(k.x)$. In this case, the solution to (\ref{KGV}) is the scalar analogue of the Volkov electron wavefunction,
\be\begin{split}\label{volkov-def-1}
	\varphi_V(x) :=  \int\!\ud\sfp\: \tilde\chi(\sfp) \varphi_\sfp(x)
	\qquad
	\varphi_\sfp(x) := \exp\bigg[-i p.x - i g_p \int\limits^{k.x}_0\! \ud s \, A_\text{ext}(s)  \bigg] \;,
\end{split}
\ee
where $p^2=m^2$ and $\tilde\chi(\sfp)$ is the (Fourier transform of) the initial data specified on a surface at $k.x=0$. The second expression in (\ref{volkov-def-1}) shows how a free plane wave is `distorted' by the presence of the external field~\cite{Lavelle:2013wx}. In analogy with previous approaches~\cite{Berson,Bergou:1980cp,Filipowicz:1985,Bagrov:1990xp,Skoromnik:2014}, we could now replace $A_\text{ext}$ in (\ref{volkov-def-1}) with a quantised field. But what would be the connection with our second quantised results? Instead, let us begin from our zero-mode theory and derive the relevant equation.

Let $\ket{\psi;x^\LCp}$ be a general solution of the Schr\"odinger equation for the zero-mode Hamiltonian (\ref{H-enkelt2-LF}), not necessarily an eigenstate. In order to connect with the first-quantised approach we project onto the one-electron sector by taking the overlap with $\bra{0}_b \phi(\sf{x})$~\cite{Bagrov:1990xp}, in which $\bra{0}_b$ is the $b$-mode vacuum. The projected state lives in the photon Fock space. Since we are interested in the effects of the three-point vertex in $H_\text{zm}$, we will also strip from these photonic states the \textit{free} time-evolution generated by the free zero-mode Hamiltonian, defining
\be\label{varphi}
	\ket{\varphi;x} := e^{ik.x a^\dagger a} \bra{0}_b\phi(\sfx) \ket{\psi;x^\LCp} \;.
\ee
Clearly knowledge of all $\ket{\psi;x^\LCp}$ is equivalent to knowledge of all $\ket{\varphi;x}$ (in the one-electron sector), so let us ask how the Schr\"odinger equation looks for the latter. Acting with $H_\text{zm}$ we find that $\ket{\varphi;x}$ obeys a \textit{light-front Schr\"odinger equation}
\be\label{LFS-1}
	i \partial_\LCp \ket{\varphi;x} = \frac{1}{4i\partial_\LCm} \bigg( - \partial_\LCperp^2 + m^2 + g A_\mathrm{zm}(k.x)\bigg) \ket{\varphi;x} \;,
\ee
in which $A_\mathrm{zm}(k.x)$ is the the scalar analogue of (\ref{K2K2}),
\be\label{AZM-DEF-NY}
	A_\mathrm{zm}(k.x) := a^\dagger e^{ik.x} + a e^{-ik.x} \;.
\ee
Rearranging (\ref{LFS-1}) gives
\be\label{KGV2}
	(\partial^2 + m^2) \ket{\varphi;x} + g A_\mathrm{zm} (k.x) \ket{\varphi;x} =0 \;,
\ee
which is a `quantised' generalisation of (\ref{KGV}), the Klein-Gordon equation, for the state $\ket{\varphi;x}$ in terms of the operator $A_\text{zm}$. This is analogous to the Dirac equation studied in the first-quantised literature approaches~\cite{Berson}, but here derived from second quantisation~\cite{Bagrov:1990xp}. (The state $\ket{\varphi;x}$ is, in the literature, typically written as $\varphi(x)$ in order to look like a wavefunction.) The general solution to (\ref{KGV2}) can be written down using the projection of the time-evolution operator in the one-electron sector, which is (see also~\cite{Mancini,Bose:1997})  
\be\label{utveckling-zm}\begin{split}
	e^{-i H_\text{zm} x^\LCp} &\to \int\!\ud\sfp\,  b^\dagger_\sfp b_\sfp
	e^{-ip_\LCp x^\LCp }
	D(-g_p) e^{-ik.x a^\dagger a} D\big(g_p) \;.
\end{split}
\ee
To make contact with the first-quantised solution (\ref{volkov-def-1}) it is convenient to expand the initial state defining the solution of the Schr\"odinger equation in a basis of eigenstates of the usual \textit{free} Hamiltonian, rather than in an eigenbasis of $H_\text{zm}$. So take the initial state to be
\be
	\int\!\ud\sfp\sum\limits_n \: \tilde\chi(\sfp,n)  b^\dagger_\sfp \ket{n} + \ldots \;,
\ee
in which the ellipses denote zero or multi-electron states which will be projected out by (\ref{varphi}).  Applying the time-evolution operator (\ref{utveckling-zm}) and the projection to (\ref{varphi}) results in
\be\begin{split}\label{volkov-def-2}
	\ket{\varphi;x} = \sum_n\int\!\ud\sfp\: \tilde\chi(\sfp,n)\: \exp\bigg[ -ip.x - i g_p \int\limits^{k.x}_0\! \ud s \, A_\text{zm}(s)\bigg] \ket{n} e^{- ig_p^2\sin(k.x)}\;,
 %
\end{split}
\ee
where we have rewritten the displacement operators $D$ in (\ref{utveckling-zm}) in terms of the operator~$A_\text{zm}$. Comparing (\ref{volkov-def-1}) and (\ref{volkov-def-2}) we immediately see a number of similarities, most notably that~$A_\text{zm}$ replaces the external field~$A_\text{ext}$ in the Volkov phase\footnote{Clearly the integral in (\ref{volkov-def-2}) can be performed, here and below, for the field (\ref{AZM-DEF-NY}), but writing it as shown makes comparison with (\ref{volkov-def-1}) direct, and also allows the extension to multiple modes with momenta all proportional to $ k_\mu$.}. However, there are additional terms in (\ref{volkov-def-2}) which arise from operator ordering when combining the displacement operators, and an extra dependency on the initial number of photons $n$. In the remainder of this section we will compare the physics in the Volkov solutions (\ref{volkov-def-1}) and the states (\ref{volkov-def-2}) in some detail.

\subsection{Coherent states and laser-particle interactions}
%
As stated above, scattering processes in a background plane wave $A_\text{ext} $ are calculated~\cite{Nikishov:1964zza,Nikishov:1964zz,Narozhnyi:1965,Nikishov:1965,landau4} using the Volkov solutions (\ref{volkov-def-1}). $A_\text{ext}$ is a prescribed field with profile chosen to model a laser, in some approximation. It can be described quantum mechanically by an initial coherent state of real photons (see below).  The one-electron states~(\ref{volkov-def-2}), including the eigenstates, are also coherent states, as shown by their dependence on the exponential of $A_\text{zm}$. This coherence is \emph{not} prescribed, but is determined by the theory as the dressing of electrons by virtual photons.

As such, if we calculate transition amplitudes (induced by some perturbation) between states of the zero-mode theory then we might expect \textit{structural} similarities with transition amplitudes calculated using the Volkov solutions~(\ref{volkov-def-1}). However, the results should not be interpreted as a `quantised' generalisation of a background laser-matter calculation, because there is no laser present, but rather a number state of photons along with a dressing of the electrons. To be concrete, we illustrate using the Compton scattering amplitude above, which we write out again for completeness:
\be\label{sfi-compton2}
\begin{split}
	S_{fi} &= ie\,(2\pi)^4 \delta^4( p' +k'- p-k) \bra{0}D(g_{p'} - g_p)\ket{1} \\
	&=  ie\,(2\pi)^4 \delta^4( p' +k' - p-k) \bigg(\frac{g}{2k.p} - \frac{g}{2k.p'}\bigg) e^{-(g_{p'}-g_p)^2/2} \;.
\end{split}
\ee
This is an ordinary transition between number states (in vacuum, i.e.~without any background field) but with all-orders quantum corrections added to account for, in a single mode approximation, the dressing of the scattered particles.  The same structure is seen in the analogous QED calculation~\cite{Bergou:1980cp}, but has been interpreted differently: due to the structural similarity with the Volkov solution, the $S$-matrix element has been identified with the process of ``nonlinear Compton scattering''~ \cite{Nikishov:1964zza,Nikishov:1964zz,Narozhnyi:1965,Nikishov:1965,landau4}, which describes the emission of a photon from an electron in the field of an intense plane wave background field, through the absorption of photons from that field. Further, the transition from $n=1 \to n=0$ physical photons is referred to as describing ``complete depletion'' of the intense mode~\cite{Bergou:1980cp}. However, there is no intense mode here, and no laser, only a single physical photon along with a dressed electron, and the only depletion is the absorption of the initial photon, as is standard in Compton scattering.

In order to investigate laser-particle scattering, we must first incorporate the laser into the asymptotic states. To do so consider an initial state which describes a physical electron together with a \textit{coherent} state of (physical) photons~\cite{Kibble:1965zza,Frantz,Gavrilov:1990qa,Ilderton:2017xbj}:
\be\label{i-def}
	\ket{i;\sfp,z} = B^\dagger_\sfp \, e^{-\frac{|z|^2}{2}}\sum\limits_{n} \frac{z^n}{n!} (A^\dagger)^n \ket{0} = 
	b^\dagger_\sfp D(-g_p) D(z) \ket{0}\;,
\ee
which contains an extra factor of $D(z)$. In contrast to the eigenstates (\ref{gen-psi-los}) of the single mode theory, $z$ here is a prescribed amplitude 
which does not depend on the electron momentum $\sfp$. Acting with the time evolution operator on the initial state $\ket{i;\sfp,z}$ we obtain the evolved state\footnote{The calculation is performed by pushing the displacement operator $D(z)$ to the left using the standard commutation result $D(\alpha) D(\beta) = D(\beta)D(\alpha) e^{\alpha \bar\beta - \bar\alpha \beta}$.}
\bea
\label{psi122}
	\ket{\psi_i;x^\LCp} &=& D(z e^{-ik.x})\, e^{i\sfp.\sfx}\varphi_\sfp(x) \, b^\dagger_\sfp D(-g_p)\ket{0} \: e^{g_p(z-\bar{z})}\;,
\eea
in which the Volkov solution $\varphi_\sfp(x) $ appears with background field
\be
	A_\mathrm{ext} = z e^{-ik.x} + \bar z e^{ik.x} \;,
\ee
and where the leading factor of $D(z e^{-ik.x})$ describes the free time-evolution of the initially prescribed coherent state\footnote{There is no $\sfx$ dependence, as this cancels between the $e^{i\sfx.\sfp}$ and $\varphi_\sfp$; we have written things in this way just to make the connection with (\ref{volkov-def-1}) clear.
Moreover the final exponential factor appears in order to recover the initial data on the hyperplane $k.x = 0$, as \eqref{volkov-def-1} is defined with an indefinite integral.}.
Due to the appearance of the Volkov solution, previous approaches have focused on recovering the background-field limit from states such as (\ref{psi122}). The argument is that for $g$ small, $z$ large, with $gz$ fixed, the photonic part of the state simplifies, reducing to the free evolution of the initial coherent state. As the photonic part is then `undisturbed' by the presence of the electrons, it is effectively a background field, the effects of which are captured by the Volkov wavefunction.

However, this may be more subtle than it first appears. Note that the representation of the state (\ref{psi122}) is not unique because we could, for example, swap the order of the $D$ operators; if we then make the approximation described above, we would not recover the entire Volkov phase. It is therefore better to examine e.g.~amplitudes, which removes the operator ordering ambiguity.

\subsection{The background field limit: nonlinear Compton scattering}\label{SECT:BGLIMIT}
We consider scattering of the initial state $\ket{i;\sfp,z}$ in (\ref{i-def}) to a final state $\ket{f;\sfp',z}\otimes\ket{k'}$ with the \textit{same} coherent state profile, an electron of momentum $\sfp'$ and, as above, a (non-zero mode) photon with momentum $k'_\mu \not= k_\mu$. In the external field limit this nonlinear Compton scattering proper, see~\cite{Nikishov:1965,landau4,Harvey:2009ry} for discussions of the relevant, i.e.~monochromatic field, case in QED. To lowest order in the interaction (\ref{V-DEF}) the scattering amplitude, call it $S_{fi}(z)$, is
\be\label{B-V-ANALOG}
	S_{fi}(z) = -i \int\!\ud x^\LCp \bra{\psi_f;x^\LCp} V \ket{\psi_i;x^\LCp} \;,
\ee
where the form of $\ket{\psi_f;x^\LCp}$ follows from (\ref{psi122}). (The emitted photon state $\ket{k'}$ evolves freely.) Now, any $D(z)$ which does \textit{not} depend on the electron momenta must commute with the interaction $V$, hence the factor $D(ze^{-ik.x})$ on the left of $\ket{\psi_i;x^\LCp}$ is removed by its inverse appearing on the right of $\bra{\psi_f;x^\LCp}$. (This is why we placed $D(z e^{-ik.x} )$ on the left of the state in (\ref{psi122}); it facilitates the comparison with the background field calculation.)  As a result the $z$-dependent displacement operators disappear~\cite{Kibble:1965zza,Frantz,Ilderton:2017xbj}, and what remains may be written in terms of the Volkov solution $\varphi_\sfp$ from (\ref{volkov-def-1}) as
\be\label{TILLBG1}
	S_{fi}(z) = -ie \int\!\ud^4 x\ \varphi^\dagger_{\sfp'}(x)e^{ik'.x} \varphi_\sfp(x)\, e^{-(g_p-g_{p'})^2/2} \;.
\ee
The exponential, coming from the dressing of the electrons, gives the same infra-red factor as earlier. The remainder of the integrand in (\ref{TILLBG1}), expressed in terms Volkov solutions, is the background field expression. 

In the limit that $gz$ is large (i.e.~the coupling $\sim eA $ of matter to the single mode coherent state is large), but $g$ itself small, the dressing factor may be approximated by unity, and $S_z$ becomes
\be\begin{split}\label{SFI-VOLKOV}
	S_{fi}(z) &\simeq 
	-i e \int\!\ud^4 x\ \varphi^\dagger_{\sfp'}(x)e^{ik'.x} \varphi_\sfp(x)  \;,
\end{split}
\ee
which is the lowest order nonlinear Compton scattering amplitude in a background monochromatic wave. Hence the background field limit is easily and unambiguously recovered at the amplitude level.

\subsection{Depletion}
Having understood the physics of the zero-mode Hamiltonian we can finally turn to depletion effects proper \cite{Seipt:2016fyu}. We have seen that a background field arises when a coherent state, say with profile $z$, is included in both the initial and final states~\cite{Kibble:1965zza,Frantz,Gavrilov:1990qa}. Following~\cite{Ilderton:2017xbj}, one way to include depletion is to allow the coherent state to change under scattering; i.e.~one should calculate amplitudes between different initial and final coherent states.

To illustrate we reconsider the nonlinear Compton amplitude from Sect.~\ref{SECT:BGLIMIT}. We take as initial and final states $\ket{i;\sfp,z_i}$ and $\ket{f;\sfp',z_f}\otimes\ket{k'}$, where now $z_f\not=z_i$. To model depletion of the field, we choose $z_f$ such that the average number of photons in the state is reduced; if, for example, we take $z_i = \lambda \in \mathbb{R}$, so that the initial (expected) number of photons is $ \lambda^2$, and $z_f = \lambda(1-\delta)$, then we can interpolate between no depletion of the coherent state at $\delta=0$ to full depletion at~$\delta=1$.  

We turn to the lowest order $S$-matrix element, $S_{fi}(z_f,z_i)$, which is expressed in terms of the time-evolved states as
\be \label{S.Z}
	S_{fi}(z_f,z_i) = -i \int\!\ud x^\LCp \bra{\textrm{f};x^\LCp} V \ket{\textrm{i};x^\LCp} \;,
\ee
and is calculated directly as above. Defining a complex valued field,
\be
	A_D(k.x) := \bar{z}_f e^{ik.x}+ z_i e^{-ik.x} \;,
\ee
which contains both the initial and final coherent state profiles, $z_i$ and $\bar{z}_f$, the $S$-matrix element (\ref{S.Z}) is found to depend on the functions~\cite{Ilderton:2017xbj}
\be\begin{split}\label{volkov-AD}
	\varphi_{p,\text{in}}(x) &:=  \exp\bigg[-i p.x - i g_p \int\limits^{k.x}\! \ud s \, A_D(s)  
	\bigg] \; ,  \\
	\varphi^\dagger_{p',\text{out}}(x) &:=  \exp\bigg[i p'.x + i g_{p'} \int\limits^{k.x}\! \ud s 
	\, A_D(s)  \bigg] \;.
\end{split}
\ee
These are identical to the (scalar Yukawa) Volkov solutions, except that $A_D$ appears in place of the external field. Using the example above, $z_i = \lambda$ and $z_f=(1-\delta)\lambda$, we have
\be
	A_D = \lambda \big(2-\delta\big)\cos k.x - i \lambda \delta \sin k.x \;,
\ee
which in the no-depletion limit, $\delta=0$, recovers the expected external field $A_\text{ext} = 2 \lambda \cos k.x$. Explicitly, in terms of (\ref{volkov-AD}), the scattering amplitude (\ref{S.Z}) is
\be\begin{split}\label{fult}
	S_{fi}(z_f,z_i)  &= 
	e^{-|z_f-z_i|^2/2 + i \text{Im}({\bar z}_f z_i)}\int\!\ud^4 x\ \varphi^\dagger_{\sfp',\text{out}}(x)e^{ik'.x} \varphi_{\sfp,\text{in}}(x)\, e^{-(g_p'-g_p)^2/2} \;.
\end{split}
\ee
The leading exponential factor accounts for the normalisation of the initial and final coherent states~\cite{Ilderton:2017xbj}, while the final exponential is the same IR factor as before. Comparing with the background field result (\ref{TILLBG1}), we see that the new functions (\ref{volkov-AD}) take the place of the Volkov solutions in the scattering amplitude. Despite this they cannot be interpreted as asymptotic single-particle wavefunctions in the sense of Volkov. Doing so would require the particles to move in a complex-valued external field $A_D$, which is hard to motivate physically. Nor can the functions be normalised, precisely because $A_D$ is complex. However, they do not need to be; they arise from a properly normalised amplitude and their form is dictated by the theory.

Pursuing this point, we comment that there is an equivalent way to write the scattering amplitude (\ref{fult}) in terms of Volkov solutions in a new external field. The real part of $A_D$ is the \emph{average} of the initial and final coherent state profiles, i.e.
\be
	A_\text{av} := \text{Re }A_D = \frac{1}{2} ({\bar z}_f + {\bar z}_i )e^{ik.x} + \text{c.c.} \;.
\ee
Let the Volkov solutions (\ref{volkov-def-1}), taken with respect to this real field, be written as $\varphi_{\sfp,\, \text{av}}$. Then the $S$-matrix element (\ref{fult}) may be expressed in terms of these functions as
\be\label{SFI-DS}
	S_{fi}(z_f,z_i) = e^{i \mathrm{Im} z_i \bar z_f} \int \ud^4 x \: e^{-\tfrac{1}{2}| (z_f-z_i) e^{-ik.x}   + g_p - g_{p'} |^2} 
\varphi^\dagger_{\sfp',\,\text{av}}(x)e^{ik'.x} \varphi_{\sfp,\,\text{av}}(x) \,.
\ee
Here the depletion of the ``external'' field $A_\text{av}$ appears through the real exponential factor in the integrand, which is typical of ``decay'' effects. We emphasise that the expressions (\ref{fult}) and (\ref{SFI-DS}) are equivalent.

The Volkov-like functions found here were predicted to arise in conjunction with depletion effects~\cite{Ilderton:2017xbj}, via LSZ reduction of $S$-matrix elements calculated between different asymptotic coherent states. Thus, our direct Hamiltonian-based calculation serves as an explicit illustration of the general approach to depletion given in~\cite{Ilderton:2017xbj}.
%

 %
\section{Conclusions} \label{SECT:CONCS}

We have performed a truncation of scalar Yukawa theory (interaction $A\phi^\dagger\phi$) which leaves the theory nontrivial but exactly solvable. This can be generalised to QED, but here we neglected gauge, spin and polarisation in order to expose the physical content, and the subtleties behind first quantised approaches discussed in the literature.

The essence of the truncation is to reduce the scalar photon $A$ to a single momentum mode. Such single mode theories have been considered in the context of intense laser-matter interactions, where the quantised single mode field is assumed to describe, fully quantum mechanically, a monochromatic laser which is usually treated as a background field. We have shown that this interpretation can be questioned; while there are many structural similarities between the background field and fully quantised approaches, the physics is different. We have reinterpreted previously obtained first-quantised results not in terms of intense fields, but in terms of the dressing of physical particles~\cite{Dirac:1955uv}. The dressing contributes exponential, infra-red-like factors to scattering amplitudes.

Indeed, intense fields already have a quantum description in terms of coherent states. We have shown that once this description is combined with our approach, then quantum corrections to the background field limit, in the form of depletion effects, are naturally included in transitions between different initial and final coherent states. Using this, we have been able to take the background field limit and seen how the usual Volkov solution reappears.

We considered three approaches to the quantisation of our single mode theory.  However neither instant form (qauntising at $t=0$) nor the usual front form (quantising at $x^\LCp=0$) yielded an explicitly exactly solvable system which allowed us to connect to existing literature results. It was only when we chose the light-front time direction and the single mode momentum direction to coincide, i.e.~when the photon was a light-front zero mode, that the theory could be solved and the connection made.

Including light-front zero modes is nontrivial; one has to quantise on hyper-surfaces involving components in two lightlike directions.  Zero modes have a reputation for being elusive, a technical complication, and possibly even an irrelevant artefact of finite volume quantisation. (Even in the first paper to realise that a light-front description of laser-particle interactions was the most natural~\cite{Neville:1971uc}, the zero modes were explicitly excluded.) However, this is not the case; zero-mode contributions can be physical~\cite{Heinzl:2003jy,Collins:2018aqt}, and have observable consequences: their importance in pair production is well known~\cite{Ji:1995ft,Tomaras:2000ag,Tomaras:2001vs}, and in some cases there can be no nonperturbative pair production without zero modes~\cite{Ilderton:2014mla}. Zero modes are also (implicitly) invoked in the guise of the plane wave laser background itself~\cite{Bakker:2013cea}. Here we have shown that going beyond the external field approximation by treating all degrees of freedom quantum mechanically, also requires zero modes to be included.

The differences between the considered approaches to quantising our theory are likely attributable to the lack of Lorentz invariance implied by the choice of a preferred photon mode direction. Physically, an instant-form zero mode describes a condensate at rest, while a light-front zero mode represents a `system' moving at the speed of light such as laser photons. This suggests that also the unspoilt space-time symmetries should be quite different in each case, corresponding to Wigner's little groups for massive and massless particles, respectively~\cite{Wigner:1939cj}. Presumably, full Lorentz invariance would be restored if the rest of the photon modes were reintroduced.

The extension of our results to multiple  co-propagating (zero) modes, or even the whole spectrum of such modes with momenta $k_\mu = \omega n_\mu$ for all $\omega >0$, is technically straightforward. Including pair-creating terms in perturbation theory would also allow us to examine corrections to vacuum polarisation effects~\cite{King:2015tba}. This would, though, give rise to more involved questions of renormalisation~\cite{Skoromnik:2015hya}, which are best addressed in QED proper.  The extension to QED is technically more involved, but we expect our interpretation to go through, as physical charges in QED are dressed~\cite{Dirac:1955uv,Bagan:1999jf,Bagan:1999jk,Ilderton:2007qy}, and the single mode interaction contributes to this dressing. This is supported by the similarity between the Yukawa and QED Compton scattering examples provided here and in~\cite{Bergou:1980cp}.

\begin{acknowledgments}
\textit{The authors thank Federico Armata for useful discussions and references. This project has received funding from the European Union's Horizon 2020 research and innovation programme under the Marie Sk\l odowska-Curie grant No.~701676 (A.I.), and the
Science and Technology Facilities Council, grant No. ST/G008248/1 (D.S.).}
\end{acknowledgments}

\appendix

\section{Instant vs. front form quantisation}
\subsection{Instant form} \label{SECT:INSTANT}

We quantise the scalar-Yukawa theory (\ref{enkel-L}) at equal time $x^0$. The Schr\"odinger picture Hamiltonian is easily written down in terms of the photon, electron and positron modes with (respective) commutators 
\be\begin{split}
 [a_{\bf p}, a^\dagger_{{\bf p}'}] =   [b_{\bf p}, b^\dagger_{{\bf p}'}] = [d_{\bf p}, d^\dagger_{{\bf p}'}]  = \delta^3({\bf p}-{\bf p}')  \;.
\end{split}\ee
We select out a single photon of momentum $\bf k$ using the two projectors
\be\label{P-Q-1}
	\mathbb{P} = \frac{\delta^3({\bf p}-{\bf k})}{\delta^3(0)} \;, \qquad \mathbb{Q}= 
	1- \mathbb{P} \;,
\ee
where some regularisation of the delta function is understood. Inserting the resolution of unity $1 = \mathbb{P} + \mathbb{Q}$ under all photon momentum integrals, any term with $\mathbb P$ is considered part of the `free' Hamiltonian $H_0$ on which we focus, while any term with $\mathbb Q$ is part of the `interacting' Hamiltonian. Retaining only $\mathbb P$-terms and rescaling the single mode operator as $a_{\bf k} = a \sqrt{\delta^3(0)}$ such that $[a,a^\dagger]=~1$, the free Hamiltonian becomes
\be\label{H00}
	H_0 = \omega_{\bf k} a^\dagger a + \int\!\ud^3{\bf p}\  E_{\bf p} (b^\dagger_{\bf p} b_{\bf p} + d^\dagger_{\bf p} d_{\bf p}) +  \frac{e}{\sqrt{2\omega_k V}}  \int\!\ud^3{\bf x}\ \phi^\dagger \phi \big( a e^{-i {\bf k}.{\bf x}} + a^\dagger e^{i {\bf k}.{\bf x}} \big) \;,
\ee
in which $E_{\bf p} = \sqrt{{\bf p}^2+m^2}$ and $\omega_{\bf k} = |\bf k|$ are the on-shell energies. The final term in $H_0$ describes all 3-point interactions between the matter fields and the single mode. This includes number changing terms such as $b^\dagger d^\dagger a^\dagger$ and $b^\dagger d^\dagger a$ (pair creation from a photon). This theory cannot be solved exactly. Noting that the existing literature uses a relativistic first quantised approach, which corresponds to \emph{fixed} fermion number, we are prompted to extend the definitions of $\mathbb P$ and $\mathbb Q$ such that matter-number changing terms are also shifted into the interaction Hamiltonian. Defining the effective coupling $g := e/(\sqrt{2\omega_k V})$~\cite{Bergou:1980cp} the new `free' Hamiltonian $H_0'$ is then 
\be\label{H-enkelt1}
	H_0' = \omega_{\bf k} a^\dagger a + \int\!\ud^3{\bf p}\  E_{\bf p} b^\dagger_{\bf p} b_{\bf p}  +  g  a \int\!\ud^3{\bf p} \frac{b^\dagger_{{\bf p}+{\bf k}} b_{\bf p}}{2\sqrt{E_{{\bf p}+{\bf k}}E_{\bf p}}} + g  a^\dagger \int\!\ud^3{\bf p} \frac{b^\dagger_{{\bf p}-{\bf k}} b_{\bf p}}{2\sqrt{E_{{\bf p}-{\bf k}}E_{\bf p}}} \;,
\ee
plus identical terms for the $d$-modes, but we do not write these explicitly as the simplifications above decouple $b$ and~$d$.

The Hamiltonian $H_0'$ describes the interaction of electrons with a single photon mode. Total momentum is conserved. The free $n$-photon number states $\ket{n}$ are eigenstates of the theory. If there are $b$-modes in the state, acting with the Hamiltonian changes photon number by $\pm 1$. Thus electron-photon eigenstates of the Hamiltonian must be infinite superpositions of multi-photon states in which the total momentum is split between the electrons and photons. To illustrate, the most general one-electron state with total momentum ${\bf p}$ is
\be\label{IF-expansion-2}
	\ket{\psi_{1e}} := \sum\limits_{n=0}^\infty  C_n \sqrt{2 E_{{\bf p}- n {\bf k}}}  \, 
	b^\dagger_{{\bf p}-n {\bf k}}\ket{n}  \;,
\ee
with normalisation coefficients $C_n$. Acting with the Hamiltonian, the Schr\"odinger equation for an eigenstate with energy~$\mathcal E$ yields a recurrence relation for the coefficients,
\be\label{Proto-RN}
	\mathcal{E} C_j = \big(j \omega + E_{{\bf p}-j{\bf k}} \big) C_j + \frac{g\sqrt{j+1}}{2 E_{{\bf p}-j{\bf k}} } C_{j+1} + \frac{g\sqrt{j}}{2 E_{{\bf p}-j{\bf k}} } C_{j-1} \;.
\ee
We have not found any closed form solution to these equations, nor any connection to the scalar Yukawa analogue of the first quantised literature solutions.

\subsection{Front form (non-zero mode)} \label{SECT:NZM}


We now quantise the scalar-Yukawa theory in the front form as reviewed in~e.g.~\cite{Burkardt:1995ct,Brodsky:1997de,Heinzl:2000ht}, defining coordinates $x^\LCpm = x^0 \pm x^3$, $x^\LCperp=(x^1,x^2)$, and taking $x^\LCp$ as the time. Momenta are $p_\LCpm  =(p_0 \pm p_3)/2$ and $p_\LCperp=\{p_1,p_2\}$. We use sans serif fonts for the light-front position and momentum three-vectors $\sfx=(x^\LCm,x^\LCperp)$ and $\sfp=(p_\LCm,p_\LCperp)$. We also define an integral measure and delta function by $\ud\sfp := \ud^2p_\LCperp \ud p_\LCm \theta(p_\LCm)$ and $\delta(\sfp - \sfq)  := \delta(p_\LCm-q_\LCm)\delta^2(p_\LCperp-q_\LCperp)$.

In the front form the Hamiltonian is
\be\label{HLF0}
	H_\text{LF} = \frac{1}{2}\int\!\ud\sfx\ \Big( \partial_\LCperp\phi^\dagger \partial_\LCperp\phi 
	+ \frac{1}{2}\partial_\LCperp A \partial_\LCperp A + e A\phi^\dagger \phi \Big) \;,
\ee
and the Schr\"odinger picture fields are expanded as 
\be\begin{split}\label{FOCK-EXP-LF}
	A(\sfx) &= \int\!\frac{\ud{\sfp}}{\sqrt{(2\pi)^32p_\LCm}}\ (e^{i \sfp.\sfx} a^\dagger_{\sfp} 
	+ e^{-i \sfp.\sfx} a_\sfp ) \;, 
	\qquad
	[a_{\sfp}, a^\dagger_{\sfp'}] = \delta(\sfp-\sfp') \;, \\
	\phi(\sfx) &= \int\!\frac{\ud\sfp}{\sqrt{(2\pi)^32p_\LCm}}\ (e^{i \sfp.\sfx} d^\dagger_\sfp 
	+ e^{-i\sfp.\sfx} b_\sfp) \;,
	\qquad
	[b_\sfp, b^\dagger_{\sfp'}] = [d_\sfp, d^\dagger_{\sfp'}] = \delta(\sfp-\sfp') \;.
\end{split}
\ee
The on-shell energies are now $p_\LCp = (p_\LCperp^2+m^2)/4p_\LCm$ and $k_\LCp = k_\LCperp^2/4k_\LCm$. Note that ``longitudinal'' momenta $p_\LCm$ are positive. Consider the interaction terms in (\ref{HLF0}) in terms of the modes (\ref{FOCK-EXP-LF}). Terms such as $a^\dagger_\sfk d^\dagger_\sfp b^\dagger_\sfq$  vanish by positivity of the longitudinal momentum since the associated delta function $\delta(\sfk+\sfp+\sfq)$ has no support. This simplification relative to the instant form is standard in the front form, but holds only if we assume that `zero modes' with vanishing longitudinal momenta are absent~\cite{Maskawa:1975ky,Heinzl:2000ht}. Otherwise, the delta function $\delta(\sfk+\sfp+\sfq)$ could be supported at $k_\LCm = p_\LCm = q_\LCm = 0$. For now we will assume $k_\LCm \ne 0$ (implying two independent directions as in Fig.~\ref{FIG:3}, middle panel) and postpone the inclusion of zero modes to the next subsection. As before, we kill number-changing terms in $H$, such as pair production from a photon, $a_\sfk b^\dagger_\sfp d^\dagger_\sfq$, by hand, truncating to the `no positron' sector. We separate out the chosen photon momentum mode  $\sfk$ by inserting into the Fourier transform of $A({\bf x})$ the resolution of unity $1 = \mathbb{P} + \mathbb{Q}$, where now~\cite{Heinzl:1991vd}
\be\label{P-Q-11}
	\mathbb{P} = \frac{\delta(\sfp-\sfk)}{\delta(\sf{0})} \;, \qquad \mathbb{Q}= 1- \mathbb{P} \;.
\ee
Doing so, and rescaling the single mode operator as $a_\sfk = a \sqrt{\delta(\sf{0})}$ with $[a,a^\dagger]=~1$, we obtain the single-mode Hamiltonian
\be\label{HLF}
	H_0 = k_\LCp a^\dagger a + \int\!\ud{\sf p}\; p_\LCp b^\dagger_\sfp b_\sfp + \frac{g}{2} \int\!\; \frac{\ud{\sfp}}{\sqrt{2p_\LCm 2(p+k)_\LCm}}\big(a b^\dagger_{\sfp+\sfk} b_\sfp + a^\dagger b^\dagger_\sfp b_{\sfp+\sfk} \big) \;,
\ee
in which $g = e/\sqrt{2k_\LCm V}$ and $V = (2\pi)^3 \delta(\sf{0})$ is the light-front volume of the perpendicular and longitudinal directions\footnote{We note for later use that the scaling above, together with the instruction to only retain $\mathbb P$ terms is equivalent to making the replacement $a_\sfp \to \delta(\sfp - \sfk )a /\sqrt{V}$.}. This is the analogue of the Hamiltonian (\ref{H-enkelt1}) in the instant form, and it appears to be very similar. However, the positivity of the longitudinal momentum gives a significant simplification when it comes to constructing the eigenstates.   The (light-front) momentum operator
\be
	\hat{\sf P} \equiv \sfk\, a^\dagger a + \int\!\ud\sfp\ \sfp \, b^\dagger_{\sfp} b_\sfp 
	 \;,
\ee
commutes with $H_0$, thus energy eigenstates can also be labelled by their conserved total momentum. The Hamiltonian again conserves electron number, and it is convenient to solve the Schr\"odinger equation sector by sector in the number of $b$-modes. (The Fock vacuum and free photon number states are eigenstates, as above.) Consider, as we did above, the one-electron sector and a state of total momentum $\sf P$. This momentum is shared between the electron and any photons present, as in (\ref{IF-expansion-2}).  As the longitudinal momentum $p_\LCm$ of the electron is positive, though, the only photon Fock states, labelled by integer $m$, which can contribute obey $p_{\LCm} = P_\LCm - m k_\LCm>0$, hence
\be
	  0 \leq m < \frac{P_\LCm}{k_\LCm}  \;,
\ee
and thus energy eigenstates must be \textit{finite} superpositions of electron momentum states\footnote{This is analogous to the situation in discretised light-cone quantisation (DLCQ), where the maximum value of $m$ is called the harmonic resolution \cite{Pauli:1985pv,Pauli:1985ps}, see also~\cite{Zhao:2013cma}.}. The allowed electron momenta in the state will obey $0 < p_\LCm \leq P_\LCm$. The most general one-electron state can now be expanded in the basis $\ket{\underline{m}}$ defined by 
\be
	\ket{\psi_{1e}} = \sum\limits_{m=0}^{\lfloor P_\LCm/k_\LCm \rfloor} 
	C_m \sqrt{2(P_\LCm - m k_\LCm)} 	\; b^\dagger_{\sfP-m \sfk}\ket{m}  
	=: \sum\limits_{m=0}^{\lfloor P_\LCm/k_\LCm \rfloor} C_m \, \ket{\underline{m}} \;.
\ee
In terms of this basis the time-independent Schr\"odinger equation for energy eigenstates reduces to a recurrence relation in analogy to (\ref{Proto-RN}). However, attempting to solve this relation directly is not the way to proceed. There is a simpler way to identify the (in terms of  the number of contributing Fock states) `low lying' eigenstates, as follows.

Suppose $P_\LCm < k_\LCm$. Then the only Fock state which can contribute to the sum is $m=0$, the vacuum. It is easily confirmed that
 \be\label{till-1}
	H_0\, \ket{\underline{0}}  = P_\LCp\, \ket{\underline{0}}\;, \quad\text{ for } P_\LCm < k_\LCm \;,
\ee
and hence a free electron is an eigenstate of the interacting theory when $P_\LCm < k_\LCm$. Now take $P_\LCm > k_\LCm$. Specifically, let $P_\LCm = nk_\LCm + \delta$ with $n\in\mathbb{Z}\geq 0$ and $0<\delta<1$ a remainder. Then the energy eigenstates are superpositions of $n+1$ terms of photon number $0\ldots n$.  The case above was $n=0$. For $n=1$, or $P_\LCm = k_\LCm + \delta$ with $0 < \delta < 1$, a general momentum eigenstate has the form 
\be
	\ket{\psi} = C_0 \, \ket{\underline 0}  +  C_1 \, \ket{\underline 1}\;,
\ee
Applying the Hamiltonian and demanding that $\ket\psi$ also be an energy eigenstate reduces the Schr\"odinger equation to a \textit{quadratic} equation for $C_1/C_0$, with two solutions
\be\label{CEN}
	\frac{C_1}{C_0} = \frac{g}{-k.P \pm \sqrt{k.P^2 + g^2 \left(\frac{P_\LCm-k_\LCm}{P_\LCm}\right)}} \;, 
\ee
which implies that the action of $H_0$ yields the eigenvalue equation
\be
	H_0\ket{\psi} = \left[P_\LCp + \frac{k.P \pm \sqrt{k.P^2 + g^2 
	\left(\frac{P_\LCm-k_\LCm}{P_\LCm}\right)}}{4 (P_\LCm -k_\LCm)}\right]\ket{\psi} \;.
\ee
We have now seen that for states of momentum ${\sf P}$ with $0< P_\LCm < k_\LCm$ there is a single eigenstate (\ref{till-1}), and for $k_\LCm < P_\LCm < 2k_\LCm$ there are two eigenstates. It is not hard to check that for $2k_\LCm < P_\LCm < 3k_\LCm$ the Schr\"odinger equation reduces to a cubic equation governing the coefficients of the mode expansion, implying (at most) three solutions, and so on. 

We comment on the limit $g\to0$. Taking the relative minus sign in (\ref{CEN}) and imposing normalisation, $C_0^2+C_1^2=1$,  we find
\be
	C_0 \to 1 \;, \quad C_1 \to 0 \;, \quad \ket{\psi}\to b_\sfP^\dagger\ket{0}  \;,\quad \text{energy} \to P_\LCp \;,
\ee
so that the state reduces to the free one-electron, no-photon state. Taking the relative plus sign in (\ref{CEN}) instead we find $C_0 \to 0$, $C_1 \to 1$, and the state reduces to the free one-electron one-photon state,
\be
	\ket{\psi}\to b_{\sfP-\sfk}^\dagger\ket{1}  \;,\quad \text{energy} \to P_\LCp + \frac{k.P}{n.(P-k)} = [P-k]_\LCp + k_\LCp\;.
\ee
In the final equality above, $[P-k]_\LCp$ is the light-front energy of an electron with momentum $\sfP-\sfk$. Thus the free field limit is correctly recovered.

We thus note a definite advantage of the front form over the instant form: we can construct some low-lying energy eigenstates explicitly due to the positivity of the longitudinal momentum $p_\LCm$. As there is no such restriction on the \textit{space-like} momenta in instant form quantisation, all eigenstates there are \textit{infinite} superpositions of photon-number states, irrespective of the total momentum of the state. However, the literature solutions to the Dirac equation with an operator-valued $A(x)$ typically involve coherent and squeezed photon states (for any given electron momentum)~\cite{Bergou:1980cm,Bergou:1980cp}. As with the instant form, then, the literature results do not seem to have anything in common with the explicit solutions (\ref{till-1}) and (\ref{CEN}) in the `non-coincident' front form.

\section{Analogous systems} \label{SECT:OPTO}
The zero-mode Hamiltonian (\ref{H-enkelt2-LF}) has a number of analogues in solid state many-body systems, in particular polarons, i.e.\ electrons dressed by phonons as modelled by the Froehlich Hamiltonian~\cite{Froehlich:1952} or similar~\cite{Lang:1963}. A more recent  analogue is optomechanics, which describes the coupling of a mirror (a mechanical, or harmonic, oscillator) to a photonic cavity mode. The associated Hamiltonian is also diagonalised by the polaron transformation of~\cite{Lang:1963}, see  \cite{Mancini,Bose:1997}. There are two differences, though. First, the roles of our fields are essentially reversed compared to those in optomechanics. Our single photon mode maps to the optomechanical mirror, while our electron modes correspond to the optomechanical photon modes. This leads to the second difference: we allow for all momentum modes of the electrons, which corresponds to allowing all frequencies of the optomechanical `cavity' EM mode. Using our notation, the optomechanical Hamiltonian would be, see e.g.~\cite{Nunnenkamp},
\be\label{H-OPTO}
	H_\text{opto} = \omega a^\dagger a + E b^\dagger b + g a^\dagger a (b + b^\dagger) \;,
\ee
in which we can see explicitly that, in comparison to (\ref{H-enkelt2-LF}) the roles of $a$ and $b$ are reversed, and that our electron/the optomechanical cavity mode has also been truncated to a single oscillator. See~\cite{Aspelmeyer} for a review and references.

We found above that considering photon zero modes leads to the automatic closure of the system as all terms in the Hamiltonian changing matter-particle number dropped out automatically. In optomechanics there are corresponding terms which could appear in (\ref{H-OPTO}), such as  $a^\dagger a^\dagger (b+b^\dagger)$; these terms are dropped under the assumptions that the cavity has a dominant mode, and that the motion of the mirror is adiabatic, such that transitions between different photon modes can be ignored~\cite{Law-Hamiltonian}. We also note that our $H_\text{zm}$ is derived from a relativistic starting point, whereas the optomechanical system is non-relativistic. The appearance of non-relativistic structures is though typical when quantising on the light-front~\cite{Brodsky:1997de,Heinzl:2000ht}.   (For a non-relativistic particle coupled to a quantised EM mode see~\cite{Bergou:1980cm,Mati:2016xdj}.)

In our dressing interpretation above we have subtracted a sector-dependent term from the Hamiltonian; we remark that whether this is reasonable may depend on the physical situation. Optomechanics is a non-relativistic system where it is natural to consider finite-time transitions between states, rather than scattering, and here there are indeed nontrivial transitions, see~\cite{Armata:2017gmp}. Our zero-mode Hamiltonian on the other hand is supposed to describe (a part of) a relativistic quantum field theory where it is natural to consider scattering, and where we expect to have to renormalise. Doing so in our case seems to make the zero-mode theory trivial \textit{in terms of scattering}, but there is still nontrivial physics in the form of dressings, which contribute to infra-red effects in perturbation theory, see (\ref{sfi-compton}).

\bibliography{single-mode-V2-bib}

\begin{thebibliography}{92}%
\makeatletter
\providecommand \@ifxundefined [1]{%
 \@ifx{#1\undefined}
}%
\providecommand \@ifnum [1]{%
 \ifnum #1\expandafter \@firstoftwo
 \else \expandafter \@secondoftwo
 \fi
}%
\providecommand \@ifx [1]{%
 \ifx #1\expandafter \@firstoftwo
 \else \expandafter \@secondoftwo
 \fi
}%
\providecommand \natexlab [1]{#1}%
\providecommand \enquote  [1]{``#1''}%
\providecommand \bibnamefont  [1]{#1}%
\providecommand \bibfnamefont [1]{#1}%
\providecommand \citenamefont [1]{#1}%
\providecommand \href@noop [0]{\@secondoftwo}%
\providecommand \href [0]{\begingroup \@sanitize@url \@href}%
\providecommand \@href[1]{\@@startlink{#1}\@@href}%
\providecommand \@@href[1]{\endgroup#1\@@endlink}%
\providecommand \@sanitize@url [0]{\catcode `\\12\catcode `\$12\catcode
  `\&12\catcode `\#12\catcode `\^12\catcode `\_12\catcode `\%12\relax}%
\providecommand \@@startlink[1]{}%
\providecommand \@@endlink[0]{}%
\providecommand \url  [0]{\begingroup\@sanitize@url \@url }%
\providecommand \@url [1]{\endgroup\@href {#1}{\urlprefix }}%
\providecommand \urlprefix  [0]{URL }%
\providecommand \Eprint [0]{\href }%
\providecommand \doibase [0]{http://dx.doi.org/}%
\providecommand \selectlanguage [0]{\@gobble}%
\providecommand \bibinfo  [0]{\@secondoftwo}%
\providecommand \bibfield  [0]{\@secondoftwo}%
\providecommand \translation [1]{[#1]}%
\providecommand \BibitemOpen [0]{}%
\providecommand \bibitemStop [0]{}%
\providecommand \bibitemNoStop [0]{.\EOS\space}%
\providecommand \EOS [0]{\spacefactor3000\relax}%
\providecommand \BibitemShut  [1]{\csname bibitem#1\endcsname}%
\let\auto@bib@innerbib\@empty
\bibitem [{\citenamefont {Belavin}\ \emph {et~al.}(1984)\citenamefont
  {Belavin}, \citenamefont {Polyakov},\ and\ \citenamefont
  {Zamolodchikov}}]{Belavin:1984vu}%
  \BibitemOpen
  \bibfield  {author} {\bibinfo {author} {\bibfnamefont {A.~A.}\ \bibnamefont
  {Belavin}}, \bibinfo {author} {\bibfnamefont {A.~M.}\ \bibnamefont
  {Polyakov}}, \ and\ \bibinfo {author} {\bibfnamefont {A.~B.}\ \bibnamefont
  {Zamolodchikov}},\ }\href {\doibase 10.1016/0550-3213(84)90052-X} {\bibfield
  {journal} {\bibinfo  {journal} {Nucl. Phys.}\ }\textbf {\bibinfo {volume}
  {B241}},\ \bibinfo {pages} {333} (\bibinfo {year} {1984})}\BibitemShut
  {NoStop}%
\bibitem [{\citenamefont {Zamolodchikov}\ and\ \citenamefont
  {Zamolodchikov}(1979)}]{Zamolodchikov:1978xm}%
  \BibitemOpen
  \bibfield  {author} {\bibinfo {author} {\bibfnamefont {A.~B.}\ \bibnamefont
  {Zamolodchikov}}\ and\ \bibinfo {author} {\bibfnamefont {A.~B.}\ \bibnamefont
  {Zamolodchikov}},\ }\href {\doibase 10.1016/0003-4916(79)90391-9} {\bibfield
  {journal} {\bibinfo  {journal} {Annals Phys.}\ }\textbf {\bibinfo {volume}
  {120}},\ \bibinfo {pages} {253} (\bibinfo {year} {1979})}\BibitemShut
  {NoStop}%
\bibitem [{\citenamefont {Bagrov}\ and\ \citenamefont
  {Gitman}(1990)}]{Bagrov:1990xp}%
  \BibitemOpen
  \bibfield  {author} {\bibinfo {author} {\bibfnamefont {V.~G.}\ \bibnamefont
  {Bagrov}}\ and\ \bibinfo {author} {\bibfnamefont {D.~M.}\ \bibnamefont
  {Gitman}},\ }\href@noop {} {\emph {\bibinfo {title} {{Exact solutions of
  relativistic wave equations}}}}\ (\bibinfo {year} {1990})\BibitemShut
  {NoStop}%
\bibitem [{\citenamefont {Heinzl}\ and\ \citenamefont
  {Ilderton}(2017)}]{Heinzl:2017zsr}%
  \BibitemOpen
  \bibfield  {author} {\bibinfo {author} {\bibfnamefont {T.}~\bibnamefont
  {Heinzl}}\ and\ \bibinfo {author} {\bibfnamefont {A.}~\bibnamefont
  {Ilderton}},\ }\href {\doibase 10.1103/PhysRevLett.118.113202} {\bibfield
  {journal} {\bibinfo  {journal} {Phys. Rev. Lett.}\ }\textbf {\bibinfo
  {volume} {118}},\ \bibinfo {pages} {113202} (\bibinfo {year} {2017})},\
  \Eprint {http://arxiv.org/abs/1701.09166} {arXiv:1701.09166 [hep-ph]}
  \BibitemShut {NoStop}%
\bibitem [{\citenamefont {Seipt}(2017)}]{Seipt:2017ckc}%
  \BibitemOpen
  \bibfield  {author} {\bibinfo {author} {\bibfnamefont {D.}~\bibnamefont
  {Seipt}},\ }in\ \href
  {https://inspirehep.net/record/1509028/files/arXiv:1701.03692.pdf} {\emph
  {\bibinfo {booktitle} {{Proceedings, QFT at the Limits: from Strong Fields to
  Heavy Quarks, (Dubna, Russia)}}}}\ (\bibinfo {year} {2017})\ pp.\ \bibinfo
  {pages} {24--43},\ \Eprint {http://arxiv.org/abs/1701.03692}
  {arXiv:1701.03692 [physics.plasm-ph]} \BibitemShut {NoStop}%
\bibitem [{\citenamefont {Yang}(2017)}]{Yang:2017xyh}%
  \BibitemOpen
  \bibfield  {author} {\bibinfo {author} {\bibfnamefont {I.-S.}\ \bibnamefont
  {Yang}},\ }\href {\doibase 10.1103/PhysRevD.96.025005} {\bibfield  {journal}
  {\bibinfo  {journal} {Phys. Rev.}\ }\textbf {\bibinfo {volume} {D96}},\
  \bibinfo {pages} {025005} (\bibinfo {year} {2017})},\ \Eprint
  {http://arxiv.org/abs/1703.03466} {arXiv:1703.03466 [hep-th]} \BibitemShut
  {NoStop}%
\bibitem [{\citenamefont {Ilderton}\ and\ \citenamefont
  {Seipt}(2018)}]{Ilderton:2017xbj}%
  \BibitemOpen
  \bibfield  {author} {\bibinfo {author} {\bibfnamefont {A.}~\bibnamefont
  {Ilderton}}\ and\ \bibinfo {author} {\bibfnamefont {D.}~\bibnamefont
  {Seipt}},\ }\href {\doibase 10.1103/PhysRevD.97.016007} {\bibfield  {journal}
  {\bibinfo  {journal} {Phys. Rev.}\ }\textbf {\bibinfo {volume} {D97}},\
  \bibinfo {pages} {016007} (\bibinfo {year} {2018})},\ \Eprint
  {http://arxiv.org/abs/1709.10085} {arXiv:1709.10085 [hep-th]} \BibitemShut
  {NoStop}%
\bibitem [{\citenamefont {Berson}(1969)}]{Berson}%
  \BibitemOpen
  \bibfield  {author} {\bibinfo {author} {\bibfnamefont {I.}~\bibnamefont
  {Berson}},\ }\href@noop {} {\bibfield  {journal} {\bibinfo  {journal} {Sov.\
  Phys.\ JETP}\ }\textbf {\bibinfo {volume} {29}},\ \bibinfo {pages} {871}
  (\bibinfo {year} {1969})}\BibitemShut {NoStop}%
\bibitem [{\citenamefont {Bergou}\ and\ \citenamefont
  {Varro}(1981)}]{Bergou:1980cp}%
  \BibitemOpen
  \bibfield  {author} {\bibinfo {author} {\bibfnamefont {J.}~\bibnamefont
  {Bergou}}\ and\ \bibinfo {author} {\bibfnamefont {S.}~\bibnamefont {Varro}},\
  }\href {\doibase 10.1088/0305-4470/14/9/023} {\bibfield  {journal} {\bibinfo
  {journal} {J. Phys.}\ }\textbf {\bibinfo {volume} {A14}},\ \bibinfo {pages}
  {2281} (\bibinfo {year} {1981})}\BibitemShut {NoStop}%
\bibitem [{\citenamefont {Filipowicz}(1985)}]{Filipowicz:1985}%
  \BibitemOpen
  \bibfield  {author} {\bibinfo {author} {\bibfnamefont {P.}~\bibnamefont
  {Filipowicz}},\ }\href@noop {} {\bibfield  {journal} {\bibinfo  {journal} {J.
  Phys. A}\ }\textbf {\bibinfo {volume} {18}},\ \bibinfo {pages} {1675}
  (\bibinfo {year} {1985})}\BibitemShut {NoStop}%
\bibitem [{\citenamefont {Skoromnik}\ and\ \citenamefont
  {Feranchuk}(2014)}]{Skoromnik:2014}%
  \BibitemOpen
  \bibfield  {author} {\bibinfo {author} {\bibfnamefont {O.}~\bibnamefont
  {Skoromnik}}\ and\ \bibinfo {author} {\bibfnamefont {I.}~\bibnamefont
  {Feranchuk}},\ }\href@noop {} {\bibfield  {journal} {\bibinfo  {journal} {J.
  Phys. B}\ }\textbf {\bibinfo {volume} {47}},\ \bibinfo {pages} {115601}
  (\bibinfo {year} {2014})}\BibitemShut {NoStop}%
\bibitem [{\citenamefont {Bogolubov}(1947)}]{Bogoliubov:1947}%
  \BibitemOpen
  \bibfield  {author} {\bibinfo {author} {\bibfnamefont {N.}~\bibnamefont
  {Bogolubov}},\ }\href@noop {} {\bibfield  {journal} {\bibinfo  {journal} {J.
  Phys. (USSR)}\ }\textbf {\bibinfo {volume} {11}},\ \bibinfo {pages} {23}
  (\bibinfo {year} {1947})}\BibitemShut {NoStop}%
\bibitem [{\citenamefont {L{\"u}scher}(1983)}]{Luscher:1982ma}%
  \BibitemOpen
  \bibfield  {author} {\bibinfo {author} {\bibfnamefont {M.}~\bibnamefont
  {L{\"u}scher}},\ }\href {\doibase 10.1016/0550-3213(83)90436-4} {\bibfield
  {journal} {\bibinfo  {journal} {Nucl. Phys.}\ }\textbf {\bibinfo {volume}
  {B219}},\ \bibinfo {pages} {233} (\bibinfo {year} {1983})}\BibitemShut
  {NoStop}%
\bibitem [{\citenamefont {L{\"u}scher}\ and\ \citenamefont
  {M{\"u}nster}(1984)}]{Luscher:1983gm}%
  \BibitemOpen
  \bibfield  {author} {\bibinfo {author} {\bibfnamefont {M.}~\bibnamefont
  {L{\"u}scher}}\ and\ \bibinfo {author} {\bibfnamefont {G.}~\bibnamefont
  {M{\"u}nster}},\ }\href {\doibase 10.1016/0550-3213(84)90038-5} {\bibfield
  {journal} {\bibinfo  {journal} {Nucl. Phys.}\ }\textbf {\bibinfo {volume}
  {B232}},\ \bibinfo {pages} {445} (\bibinfo {year} {1984})}\BibitemShut
  {NoStop}%
\bibitem [{\citenamefont {Matinyan}\ \emph {et~al.}(1981)\citenamefont
  {Matinyan}, \citenamefont {Savvidy},\ and\ \citenamefont
  {Ter-Arutunian~Savvidy}}]{Matinyan:1981dj}%
  \BibitemOpen
  \bibfield  {author} {\bibinfo {author} {\bibfnamefont {S.~G.}\ \bibnamefont
  {Matinyan}}, \bibinfo {author} {\bibfnamefont {G.~K.}\ \bibnamefont
  {Savvidy}}, \ and\ \bibinfo {author} {\bibfnamefont {N.~G.}\ \bibnamefont
  {Ter-Arutunian~Savvidy}},\ }\href@noop {} {\bibfield  {journal} {\bibinfo
  {journal} {Sov. Phys. JETP}\ }\textbf {\bibinfo {volume} {53}},\ \bibinfo
  {pages} {421} (\bibinfo {year} {1981})},\ \bibinfo {note} {[Zh. Eksp. Teor.
  Fiz. 80, 830(1981)]}\BibitemShut {NoStop}%
\bibitem [{\citenamefont {Nikolaevskii}\ and\ \citenamefont
  {Shur}(1983)}]{Nikolaevskii:1983}%
  \BibitemOpen
  \bibfield  {author} {\bibinfo {author} {\bibfnamefont {E.}~\bibnamefont
  {Nikolaevskii}}\ and\ \bibinfo {author} {\bibfnamefont {L.}~\bibnamefont
  {Shur}},\ }\href@noop {} {\bibfield  {journal} {\bibinfo  {journal} {JETP
  Lett.}\ }\textbf {\bibinfo {volume} {36}},\ \bibinfo {pages} {218} (\bibinfo
  {year} {1983})}\BibitemShut {NoStop}%
\bibitem [{\citenamefont {Biro}\ \emph {et~al.}(1994)\citenamefont {Biro},
  \citenamefont {Matinyan},\ and\ \citenamefont {M{\"u}ller}}]{Biro:1994bi}%
  \BibitemOpen
  \bibfield  {author} {\bibinfo {author} {\bibfnamefont {T.~S.}\ \bibnamefont
  {Biro}}, \bibinfo {author} {\bibfnamefont {S.~G.}\ \bibnamefont {Matinyan}},
  \ and\ \bibinfo {author} {\bibfnamefont {B.}~\bibnamefont {M{\"u}ller}},\
  }\href@noop {} {\bibfield  {journal} {\bibinfo  {journal} {World Sci. Lect.
  Notes Phys.}\ }\textbf {\bibinfo {volume} {56}},\ \bibinfo {pages} {1}
  (\bibinfo {year} {1994})}\BibitemShut {NoStop}%
\bibitem [{\citenamefont {Wolkow}(1935)}]{Volkov}%
  \BibitemOpen
  \bibfield  {author} {\bibinfo {author} {\bibfnamefont {D.~M.}\ \bibnamefont
  {Wolkow}},\ }\href {\doibase 10.1007/BF01331022} {\bibfield  {journal}
  {\bibinfo  {journal} {Z. Phys.}\ }\textbf {\bibinfo {volume} {94}},\ \bibinfo
  {pages} {250} (\bibinfo {year} {1935})}\BibitemShut {NoStop}%
\bibitem [{\citenamefont {Furry}(1951)}]{Furry:1951zz}%
  \BibitemOpen
  \bibfield  {author} {\bibinfo {author} {\bibfnamefont {W.~H.}\ \bibnamefont
  {Furry}},\ }\href {\doibase 10.1103/PhysRev.81.915} {\bibfield  {journal}
  {\bibinfo  {journal} {Phys. Rev.}\ }\textbf {\bibinfo {volume} {81}},\
  \bibinfo {pages} {115} (\bibinfo {year} {1951})}\BibitemShut {NoStop}%
\bibitem [{\citenamefont {Fradkin}\ and\ \citenamefont
  {Gitman}(1981)}]{Fradkin:1981sc}%
  \BibitemOpen
  \bibfield  {author} {\bibinfo {author} {\bibfnamefont {E.~S.}\ \bibnamefont
  {Fradkin}}\ and\ \bibinfo {author} {\bibfnamefont {D.~M.}\ \bibnamefont
  {Gitman}},\ }\href {\doibase 10.1002/prop.19810290902} {\bibfield  {journal}
  {\bibinfo  {journal} {Fortsch. Phys.}\ }\textbf {\bibinfo {volume} {29}},\
  \bibinfo {pages} {381} (\bibinfo {year} {1981})}\BibitemShut {NoStop}%
\bibitem [{\citenamefont {Ilderton}\ and\ \citenamefont
  {Torgrimsson}(2013)}]{Ilderton:2012qe}%
  \BibitemOpen
  \bibfield  {author} {\bibinfo {author} {\bibfnamefont {A.}~\bibnamefont
  {Ilderton}}\ and\ \bibinfo {author} {\bibfnamefont {G.}~\bibnamefont
  {Torgrimsson}},\ }\href {\doibase 10.1103/PhysRevD.87.085040} {\bibfield
  {journal} {\bibinfo  {journal} {Phys. Rev.}\ }\textbf {\bibinfo {volume}
  {D87}},\ \bibinfo {pages} {085040} (\bibinfo {year} {2013})},\ \Eprint
  {http://arxiv.org/abs/1210.6840} {arXiv:1210.6840 [hep-th]} \BibitemShut
  {NoStop}%
\bibitem [{\citenamefont {Lavelle}\ and\ \citenamefont
  {McMullan}(2015)}]{Lavelle:2015jxa}%
  \BibitemOpen
  \bibfield  {author} {\bibinfo {author} {\bibfnamefont {M.}~\bibnamefont
  {Lavelle}}\ and\ \bibinfo {author} {\bibfnamefont {D.}~\bibnamefont
  {McMullan}},\ }\href {\doibase 10.1103/PhysRevD.91.105022} {\bibfield
  {journal} {\bibinfo  {journal} {Phys. Rev.}\ }\textbf {\bibinfo {volume}
  {D91}},\ \bibinfo {pages} {105022} (\bibinfo {year} {2015})},\ \Eprint
  {http://arxiv.org/abs/1502.06529} {arXiv:1502.06529 [hep-ph]} \BibitemShut
  {NoStop}%
\bibitem [{\citenamefont {Di~Piazza}(2018)}]{DiPiazza:2018ofz}%
  \BibitemOpen
  \bibfield  {author} {\bibinfo {author} {\bibfnamefont {A.}~\bibnamefont
  {Di~Piazza}},\ }\href@noop {} {\  (\bibinfo {year} {2018})},\ \Eprint
  {http://arxiv.org/abs/1802.03202} {arXiv:1802.03202 [hep-ph]} \BibitemShut
  {NoStop}%
\bibitem [{\citenamefont {Kibble}(1965)}]{Kibble:1965zza}%
  \BibitemOpen
  \bibfield  {author} {\bibinfo {author} {\bibfnamefont {T.~W.~B.}\
  \bibnamefont {Kibble}},\ }\href {\doibase 10.1103/PhysRev.138.B740}
  {\bibfield  {journal} {\bibinfo  {journal} {Phys. Rev.}\ }\textbf {\bibinfo
  {volume} {138}},\ \bibinfo {pages} {B740} (\bibinfo {year}
  {1965})}\BibitemShut {NoStop}%
\bibitem [{\citenamefont {Frantz}(1965)}]{Frantz}%
  \BibitemOpen
  \bibfield  {author} {\bibinfo {author} {\bibfnamefont {L.~M.}\ \bibnamefont
  {Frantz}},\ }\href {\doibase 10.1103/PhysRev.139.B1326} {\bibfield  {journal}
  {\bibinfo  {journal} {Phys. Rev.}\ }\textbf {\bibinfo {volume} {139}},\
  \bibinfo {pages} {B1326} (\bibinfo {year} {1965})}\BibitemShut {NoStop}%
\bibitem [{\citenamefont {Gavrilov}\ and\ \citenamefont
  {Gitman}(1990)}]{Gavrilov:1990qa}%
  \BibitemOpen
  \bibfield  {author} {\bibinfo {author} {\bibfnamefont {S.~P.}\ \bibnamefont
  {Gavrilov}}\ and\ \bibinfo {author} {\bibfnamefont {D.~M.}\ \bibnamefont
  {Gitman}},\ }\href@noop {} {\bibfield  {journal} {\bibinfo  {journal} {Sov.
  J. Nucl. Phys.}\ }\textbf {\bibinfo {volume} {51}},\ \bibinfo {pages} {1040}
  (\bibinfo {year} {1990})},\ \bibinfo {note} {[Yad.
  Fiz.51,1644(1990)]}\BibitemShut {NoStop}%
\bibitem [{\citenamefont {Nelson}(1964)}]{Nelson:1964}%
  \BibitemOpen
  \bibfield  {author} {\bibinfo {author} {\bibfnamefont {E.}~\bibnamefont
  {Nelson}},\ }\href@noop {} {\bibfield  {journal} {\bibinfo  {journal} {J.
  Math. Phys.}\ }\textbf {\bibinfo {volume} {5}},\ \bibinfo {pages} {1190}
  (\bibinfo {year} {1964})}\BibitemShut {NoStop}%
\bibitem [{\citenamefont {Dybalski}(2017)}]{Dybalski:2017mip}%
  \BibitemOpen
  \bibfield  {author} {\bibinfo {author} {\bibfnamefont {W.}~\bibnamefont
  {Dybalski}},\ }\href {\doibase 10.1016/j.nuclphysb.2017.10.018} {\bibfield
  {journal} {\bibinfo  {journal} {Nucl. Phys.}\ }\textbf {\bibinfo {volume}
  {B925}},\ \bibinfo {pages} {455} (\bibinfo {year} {2017})},\ \Eprint
  {http://arxiv.org/abs/1706.09057} {arXiv:1706.09057 [hep-th]} \BibitemShut
  {NoStop}%
\bibitem [{\citenamefont {Dietrich}\ \emph {et~al.}(2007)\citenamefont
  {Dietrich}, \citenamefont {Hoyer}, \citenamefont {Jarvinen},\ and\
  \citenamefont {Peigne}}]{Dietrich:2006fw}%
  \BibitemOpen
  \bibfield  {author} {\bibinfo {author} {\bibfnamefont {D.~D.}\ \bibnamefont
  {Dietrich}}, \bibinfo {author} {\bibfnamefont {P.}~\bibnamefont {Hoyer}},
  \bibinfo {author} {\bibfnamefont {M.}~\bibnamefont {Jarvinen}}, \ and\
  \bibinfo {author} {\bibfnamefont {S.}~\bibnamefont {Peigne}},\ }\href
  {\doibase 10.1088/1126-6708/2007/03/105} {\bibfield  {journal} {\bibinfo
  {journal} {JHEP}\ }\textbf {\bibinfo {volume} {03}},\ \bibinfo {pages} {105}
  (\bibinfo {year} {2007})},\ \Eprint {http://arxiv.org/abs/hep-ph/0608075}
  {arXiv:hep-ph/0608075 [hep-ph]} \BibitemShut {NoStop}%
\bibitem [{\citenamefont {Dirac}(1949)}]{Dirac:1949cp}%
  \BibitemOpen
  \bibfield  {author} {\bibinfo {author} {\bibfnamefont {P.~A.~M.}\
  \bibnamefont {Dirac}},\ }\href {\doibase 10.1103/RevModPhys.21.392}
  {\bibfield  {journal} {\bibinfo  {journal} {Rev. Mod. Phys.}\ }\textbf
  {\bibinfo {volume} {21}},\ \bibinfo {pages} {392} (\bibinfo {year}
  {1949})}\BibitemShut {NoStop}%
\bibitem [{\citenamefont {Burkardt}(1996)}]{Burkardt:1995ct}%
  \BibitemOpen
  \bibfield  {author} {\bibinfo {author} {\bibfnamefont {M.}~\bibnamefont
  {Burkardt}},\ }\href {\doibase 10.1007/0-306-47067-5_1} {\bibfield  {journal}
  {\bibinfo  {journal} {Adv. Nucl. Phys.}\ }\textbf {\bibinfo {volume} {23}},\
  \bibinfo {pages} {1} (\bibinfo {year} {1996})},\ \Eprint
  {http://arxiv.org/abs/hep-ph/9505259} {arXiv:hep-ph/9505259 [hep-ph]}
  \BibitemShut {NoStop}%
\bibitem [{\citenamefont {Brodsky}\ \emph {et~al.}(1998)\citenamefont
  {Brodsky}, \citenamefont {Pauli},\ and\ \citenamefont
  {Pinsky}}]{Brodsky:1997de}%
  \BibitemOpen
  \bibfield  {author} {\bibinfo {author} {\bibfnamefont {S.~J.}\ \bibnamefont
  {Brodsky}}, \bibinfo {author} {\bibfnamefont {H.-C.}\ \bibnamefont {Pauli}},
  \ and\ \bibinfo {author} {\bibfnamefont {S.~S.}\ \bibnamefont {Pinsky}},\
  }\href {\doibase 10.1016/S0370-1573(97)00089-6} {\bibfield  {journal}
  {\bibinfo  {journal} {Phys. Rept.}\ }\textbf {\bibinfo {volume} {301}},\
  \bibinfo {pages} {299} (\bibinfo {year} {1998})},\ \Eprint
  {http://arxiv.org/abs/hep-ph/9705477} {arXiv:hep-ph/9705477 [hep-ph]}
  \BibitemShut {NoStop}%
\bibitem [{\citenamefont {Heinzl}(2001)}]{Heinzl:2000ht}%
  \BibitemOpen
  \bibfield  {author} {\bibinfo {author} {\bibfnamefont {T.}~\bibnamefont
  {Heinzl}},\ }\bibfield  {booktitle} {\emph {\bibinfo {booktitle} {{Methods of
  quantization}}},\ }\href {\doibase 10.1007/3-540-45114-5_2} {\bibfield
  {journal} {\bibinfo  {journal} {Lect. Notes Phys.}\ }\textbf {\bibinfo
  {volume} {572}},\ \bibinfo {pages} {55} (\bibinfo {year} {2001})},\ \Eprint
  {http://arxiv.org/abs/hep-th/0008096} {arXiv:hep-th/0008096 [hep-th]}
  \BibitemShut {NoStop}%
\bibitem [{\citenamefont {Heinzl}(2003)}]{Heinzl:2003jy}%
  \BibitemOpen
  \bibfield  {author} {\bibinfo {author} {\bibfnamefont {T.}~\bibnamefont
  {Heinzl}}\ }(\bibinfo {year} {2003})\ \Eprint
  {http://arxiv.org/abs/hep-th/0310165} {arXiv:hep-th/0310165 [hep-th]}
  \BibitemShut {NoStop}%
\bibitem [{\citenamefont {Collins}(2018)}]{Collins:2018aqt}%
  \BibitemOpen
  \bibfield  {author} {\bibinfo {author} {\bibfnamefont {J.}~\bibnamefont
  {Collins}},\ }\href@noop {} {\  (\bibinfo {year} {2018})},\ \Eprint
  {http://arxiv.org/abs/1801.03960} {arXiv:1801.03960 [hep-ph]} \BibitemShut
  {NoStop}%
\bibitem [{\citenamefont {Schwinger}(1962)}]{Schwinger:1962tn}%
  \BibitemOpen
  \bibfield  {author} {\bibinfo {author} {\bibfnamefont {J.~S.}\ \bibnamefont
  {Schwinger}},\ }\href {\doibase 10.1103/PhysRev.125.397} {\bibfield
  {journal} {\bibinfo  {journal} {Phys. Rev.}\ }\textbf {\bibinfo {volume}
  {125}},\ \bibinfo {pages} {397} (\bibinfo {year} {1962})}\BibitemShut
  {NoStop}%
\bibitem [{\citenamefont {McCartor}(1988)}]{McCartor:1988bc}%
  \BibitemOpen
  \bibfield  {author} {\bibinfo {author} {\bibfnamefont {G.}~\bibnamefont
  {McCartor}},\ }\href {\doibase 10.1007/BF01566926} {\bibfield  {journal}
  {\bibinfo  {journal} {Z. Phys.}\ }\textbf {\bibinfo {volume} {C41}},\
  \bibinfo {pages} {271} (\bibinfo {year} {1988})}\BibitemShut {NoStop}%
\bibitem [{\citenamefont {Hornbostel}(1988)}]{diss:Hornbostel}%
  \BibitemOpen
  \bibfield  {author} {\bibinfo {author} {\bibfnamefont {K.~J.}\ \bibnamefont
  {Hornbostel}},\ }\emph {\bibinfo {title} {The Application of Light-Cone
  Quantization to Quantum Chromodynamics in 1+1 Dimensions}},\ \href@noop {}
  {\bibinfo {type} {{PhD} thesis}},\ \bibinfo  {school} {Stanford University}
  (\bibinfo {year} {1988})\BibitemShut {NoStop}%
\bibitem [{\citenamefont {Bogolubov}\ \emph {et~al.}(1990)\citenamefont
  {Bogolubov}, \citenamefont {Logunov}, \citenamefont {Oksak},\ and\
  \citenamefont {Todorov}}]{Bogoliubov:1990}%
  \BibitemOpen
  \bibfield  {author} {\bibinfo {author} {\bibfnamefont {N.}~\bibnamefont
  {Bogolubov}}, \bibinfo {author} {\bibfnamefont {A.}~\bibnamefont {Logunov}},
  \bibinfo {author} {\bibfnamefont {A.}~\bibnamefont {Oksak}}, \ and\ \bibinfo
  {author} {\bibfnamefont {I.}~\bibnamefont {Todorov}},\ }\href@noop {} {\emph
  {\bibinfo {title} {{General Principles of Quantum Field Theory}}}}\ (\bibinfo
   {publisher} {Kluwer Academic, Dordrecht},\ \bibinfo {year}
  {1990})\BibitemShut {NoStop}%
\bibitem [{\citenamefont {Heinzl}\ \emph {et~al.}(1991)\citenamefont {Heinzl},
  \citenamefont {Krusche},\ and\ \citenamefont {Werner}}]{Heinzl:1991vd}%
  \BibitemOpen
  \bibfield  {author} {\bibinfo {author} {\bibfnamefont {T.}~\bibnamefont
  {Heinzl}}, \bibinfo {author} {\bibfnamefont {S.}~\bibnamefont {Krusche}}, \
  and\ \bibinfo {author} {\bibfnamefont {E.}~\bibnamefont {Werner}},\ }\href
  {\doibase 10.1016/0370-2693(91)90218-F} {\bibfield  {journal} {\bibinfo
  {journal} {Phys. Lett.}\ }\textbf {\bibinfo {volume} {B256}},\ \bibinfo
  {pages} {55} (\bibinfo {year} {1991})}\BibitemShut {NoStop}%
\bibitem [{\citenamefont {Maskawa}\ and\ \citenamefont
  {Yamawaki}(1976)}]{Maskawa:1975ky}%
  \BibitemOpen
  \bibfield  {author} {\bibinfo {author} {\bibfnamefont {T.}~\bibnamefont
  {Maskawa}}\ and\ \bibinfo {author} {\bibfnamefont {K.}~\bibnamefont
  {Yamawaki}},\ }\href {\doibase 10.1143/PTP.56.270} {\bibfield  {journal}
  {\bibinfo  {journal} {Prog. Theor. Phys.}\ }\textbf {\bibinfo {volume}
  {56}},\ \bibinfo {pages} {270} (\bibinfo {year} {1976})}\BibitemShut
  {NoStop}%
\bibitem [{\citenamefont {Lang}\ and\ \citenamefont
  {Firsov}(1963)}]{Lang:1963}%
  \BibitemOpen
  \bibfield  {author} {\bibinfo {author} {\bibfnamefont {I.}~\bibnamefont
  {Lang}}\ and\ \bibinfo {author} {\bibfnamefont {Y.~A.}\ \bibnamefont
  {Firsov}},\ }\href@noop {} {\bibfield  {journal} {\bibinfo  {journal}
  {Sov.~Phys.~JETP}\ }\textbf {\bibinfo {volume} {16}},\ \bibinfo {pages}
  {1301} (\bibinfo {year} {1963})}\BibitemShut {NoStop}%
\bibitem [{\citenamefont {Mancini}\ \emph {et~al.}(1997)\citenamefont
  {Mancini}, \citenamefont {Man'ko},\ and\ \citenamefont {Tombesi}}]{Mancini}%
  \BibitemOpen
  \bibfield  {author} {\bibinfo {author} {\bibfnamefont {S.}~\bibnamefont
  {Mancini}}, \bibinfo {author} {\bibfnamefont {V.~I.}\ \bibnamefont {Man'ko}},
  \ and\ \bibinfo {author} {\bibfnamefont {P.}~\bibnamefont {Tombesi}},\ }\href
  {\doibase 10.1103/PhysRevA.55.3042} {\bibfield  {journal} {\bibinfo
  {journal} {Phys. Rev. A}\ }\textbf {\bibinfo {volume} {55}},\ \bibinfo
  {pages} {3042} (\bibinfo {year} {1997})}\BibitemShut {NoStop}%
\bibitem [{\citenamefont {Bose}\ \emph {et~al.}(1997)\citenamefont {Bose},
  \citenamefont {Jacobs},\ and\ \citenamefont {Knight}}]{Bose:1997}%
  \BibitemOpen
  \bibfield  {author} {\bibinfo {author} {\bibfnamefont {S.}~\bibnamefont
  {Bose}}, \bibinfo {author} {\bibfnamefont {K.}~\bibnamefont {Jacobs}}, \ and\
  \bibinfo {author} {\bibfnamefont {P.~L.}\ \bibnamefont {Knight}},\ }\href
  {\doibase 10.1103/PhysRevA.56.4175} {\bibfield  {journal} {\bibinfo
  {journal} {Phys. Rev. A}\ }\textbf {\bibinfo {volume} {56}},\ \bibinfo
  {pages} {4175} (\bibinfo {year} {1997})}\BibitemShut {NoStop}%
\bibitem [{\citenamefont {Aspelmeyer}\ \emph {et~al.}(2014)\citenamefont
  {Aspelmeyer}, \citenamefont {Kippenberg},\ and\ \citenamefont
  {Marquardt}}]{Aspelmeyer}%
  \BibitemOpen
  \bibfield  {author} {\bibinfo {author} {\bibfnamefont {M.}~\bibnamefont
  {Aspelmeyer}}, \bibinfo {author} {\bibfnamefont {T.~J.}\ \bibnamefont
  {Kippenberg}}, \ and\ \bibinfo {author} {\bibfnamefont {F.}~\bibnamefont
  {Marquardt}},\ }\href {\doibase 10.1103/RevModPhys.86.1391} {\bibfield
  {journal} {\bibinfo  {journal} {Rev. Mod. Phys.}\ }\textbf {\bibinfo {volume}
  {86}},\ \bibinfo {pages} {1391} (\bibinfo {year} {2014})}\BibitemShut
  {NoStop}%
\bibitem [{\citenamefont {Boiteux}\ and\ \citenamefont
  {Levelut}(1973)}]{Boiteux:1973}%
  \BibitemOpen
  \bibfield  {author} {\bibinfo {author} {\bibfnamefont {M.}~\bibnamefont
  {Boiteux}}\ and\ \bibinfo {author} {\bibfnamefont {A.}~\bibnamefont
  {Levelut}},\ }\href {http://stacks.iop.org/0301-0015/6/i=5/a=004} {\bibfield
  {journal} {\bibinfo  {journal} {J. Phys. A}\ }\textbf {\bibinfo {volume}
  {6}},\ \bibinfo {pages} {589} (\bibinfo {year} {1973})}\BibitemShut {NoStop}%
\bibitem [{\citenamefont {W{\"u}nsche}(1991)}]{Wuensche:1991}%
  \BibitemOpen
  \bibfield  {author} {\bibinfo {author} {\bibfnamefont {A.}~\bibnamefont
  {W{\"u}nsche}},\ }\href@noop {} {\bibfield  {journal} {\bibinfo  {journal}
  {Quantum Opt.}\ }\textbf {\bibinfo {volume} {3}},\ \bibinfo {pages} {359}
  (\bibinfo {year} {1991})}\BibitemShut {NoStop}%
\bibitem [{\citenamefont {Henley}\ and\ \citenamefont
  {Thirring}(1962)}]{Henley:1962}%
  \BibitemOpen
  \bibfield  {author} {\bibinfo {author} {\bibfnamefont {E.}~\bibnamefont
  {Henley}}\ and\ \bibinfo {author} {\bibfnamefont {W.}~\bibnamefont
  {Thirring}},\ }\href@noop {} {\emph {\bibinfo {title} {{Elementary quantum
  field theory}}}}\ (\bibinfo  {publisher} {McGraw-Hill, New York},\ \bibinfo
  {year} {1962})\BibitemShut {NoStop}%
\bibitem [{\citenamefont {Compagno}\ \emph {et~al.}(1995)\citenamefont
  {Compagno}, \citenamefont {Passante},\ and\ \citenamefont
  {Persico}}]{Compagno:1995}%
  \BibitemOpen
  \bibfield  {author} {\bibinfo {author} {\bibfnamefont {G.}~\bibnamefont
  {Compagno}}, \bibinfo {author} {\bibfnamefont {R.}~\bibnamefont {Passante}},
  \ and\ \bibinfo {author} {\bibfnamefont {F.}~\bibnamefont {Persico}},\
  }\href@noop {} {\emph {\bibinfo {title} {{Atom-Field Interactions and Dressed
  Atoms}}}}\ (\bibinfo  {publisher} {Cambridge University Press, Cambridge,
  UK},\ \bibinfo {year} {1995})\BibitemShut {NoStop}%
\bibitem [{\citenamefont {Dirac}(1955)}]{Dirac:1955uv}%
  \BibitemOpen
  \bibfield  {author} {\bibinfo {author} {\bibfnamefont {P.~A.~M.}\
  \bibnamefont {Dirac}},\ }\href {\doibase 10.1139/p55-081} {\bibfield
  {journal} {\bibinfo  {journal} {Can. J. Phys.}\ }\textbf {\bibinfo {volume}
  {33}},\ \bibinfo {pages} {650} (\bibinfo {year} {1955})}\BibitemShut
  {NoStop}%
\bibitem [{\citenamefont {Bagan}\ \emph
  {et~al.}(2000{\natexlab{a}})\citenamefont {Bagan}, \citenamefont {Lavelle},\
  and\ \citenamefont {McMullan}}]{Bagan:1999jf}%
  \BibitemOpen
  \bibfield  {author} {\bibinfo {author} {\bibfnamefont {E.}~\bibnamefont
  {Bagan}}, \bibinfo {author} {\bibfnamefont {M.}~\bibnamefont {Lavelle}}, \
  and\ \bibinfo {author} {\bibfnamefont {D.}~\bibnamefont {McMullan}},\ }\href
  {\doibase 10.1006/aphy.2000.6048} {\bibfield  {journal} {\bibinfo  {journal}
  {Annals Phys.}\ }\textbf {\bibinfo {volume} {282}},\ \bibinfo {pages} {471}
  (\bibinfo {year} {2000}{\natexlab{a}})},\ \Eprint
  {http://arxiv.org/abs/hep-ph/9909257} {arXiv:hep-ph/9909257 [hep-ph]}
  \BibitemShut {NoStop}%
\bibitem [{\citenamefont {Bagan}\ \emph
  {et~al.}(2000{\natexlab{b}})\citenamefont {Bagan}, \citenamefont {Lavelle},\
  and\ \citenamefont {McMullan}}]{Bagan:1999jk}%
  \BibitemOpen
  \bibfield  {author} {\bibinfo {author} {\bibfnamefont {E.}~\bibnamefont
  {Bagan}}, \bibinfo {author} {\bibfnamefont {M.}~\bibnamefont {Lavelle}}, \
  and\ \bibinfo {author} {\bibfnamefont {D.}~\bibnamefont {McMullan}},\ }\href
  {\doibase 10.1006/aphy.2000.6049} {\bibfield  {journal} {\bibinfo  {journal}
  {Annals Phys.}\ }\textbf {\bibinfo {volume} {282}},\ \bibinfo {pages} {503}
  (\bibinfo {year} {2000}{\natexlab{b}})},\ \Eprint
  {http://arxiv.org/abs/hep-ph/9909262} {arXiv:hep-ph/9909262 [hep-ph]}
  \BibitemShut {NoStop}%
\bibitem [{\citenamefont {Ilderton}\ \emph {et~al.}(2007)\citenamefont
  {Ilderton}, \citenamefont {Lavelle},\ and\ \citenamefont
  {McMullan}}]{Ilderton:2007qy}%
  \BibitemOpen
  \bibfield  {author} {\bibinfo {author} {\bibfnamefont {A.}~\bibnamefont
  {Ilderton}}, \bibinfo {author} {\bibfnamefont {M.}~\bibnamefont {Lavelle}}, \
  and\ \bibinfo {author} {\bibfnamefont {D.}~\bibnamefont {McMullan}},\ }\href
  {\doibase 10.1088/1126-6708/2007/03/044} {\bibfield  {journal} {\bibinfo
  {journal} {JHEP}\ }\textbf {\bibinfo {volume} {03}},\ \bibinfo {pages} {044}
  (\bibinfo {year} {2007})},\ \Eprint {http://arxiv.org/abs/hep-th/0701168}
  {arXiv:hep-th/0701168 [hep-th]} \BibitemShut {NoStop}%
\bibitem [{\citenamefont {Lee}(1981)}]{Lee:1981mf}%
  \BibitemOpen
  \bibfield  {author} {\bibinfo {author} {\bibfnamefont {T.~D.}\ \bibnamefont
  {Lee}},\ }\href@noop {} {\emph {\bibinfo {title} {{Particle Physics and
  Introduction to Field Theory}}}}\ (\bibinfo  {publisher} {Harwood Academic},\
  \bibinfo {address} {Chur},\ \bibinfo {year} {1981})\BibitemShut {NoStop}%
\bibitem [{\citenamefont {Feynman}(1950)}]{Feynman:1950ir}%
  \BibitemOpen
  \bibfield  {author} {\bibinfo {author} {\bibfnamefont {R.~P.}\ \bibnamefont
  {Feynman}},\ }\href {\doibase 10.1103/PhysRev.80.440} {\bibfield  {journal}
  {\bibinfo  {journal} {Phys. Rev.}\ }\textbf {\bibinfo {volume} {80}},\
  \bibinfo {pages} {440} (\bibinfo {year} {1950})}\BibitemShut {NoStop}%
\bibitem [{\citenamefont {Cahill}\ and\ \citenamefont
  {Glauber}(1969)}]{Cahill:1969it}%
  \BibitemOpen
  \bibfield  {author} {\bibinfo {author} {\bibfnamefont {K.~E.}\ \bibnamefont
  {Cahill}}\ and\ \bibinfo {author} {\bibfnamefont {R.~J.}\ \bibnamefont
  {Glauber}},\ }\href {\doibase 10.1103/PhysRev.177.1857} {\bibfield  {journal}
  {\bibinfo  {journal} {Phys. Rev.}\ }\textbf {\bibinfo {volume} {177}},\
  \bibinfo {pages} {1857} (\bibinfo {year} {1969})}\BibitemShut {NoStop}%
\bibitem [{\citenamefont {Bialynicki-Birula}\ and\ \citenamefont
  {Bialynicka-Birula}(1973)}]{BB}%
  \BibitemOpen
  \bibfield  {author} {\bibinfo {author} {\bibfnamefont {I.}~\bibnamefont
  {Bialynicki-Birula}}\ and\ \bibinfo {author} {\bibfnamefont {Z.}~\bibnamefont
  {Bialynicka-Birula}},\ }\href {\doibase 10.1103/PhysRevA.8.3146} {\bibfield
  {journal} {\bibinfo  {journal} {Phys. Rev.}\ }\textbf {\bibinfo {volume}
  {A8}},\ \bibinfo {pages} {3146} (\bibinfo {year} {1973})}\BibitemShut
  {NoStop}%
\bibitem [{\citenamefont {Schwinger}\ and\ \citenamefont
  {Tsai}(1978)}]{Schwinger:1977ba}%
  \BibitemOpen
  \bibfield  {author} {\bibinfo {author} {\bibfnamefont {J.~S.}\ \bibnamefont
  {Schwinger}}\ and\ \bibinfo {author} {\bibfnamefont {W.-y.}\ \bibnamefont
  {Tsai}},\ }\href {\doibase 10.1016/0003-4916(78)90142-2} {\bibfield
  {journal} {\bibinfo  {journal} {Annals Phys.}\ }\textbf {\bibinfo {volume}
  {110}},\ \bibinfo {pages} {63} (\bibinfo {year} {1978})}\BibitemShut
  {NoStop}%
\bibitem [{\citenamefont {Chung}(1965)}]{Chung:1965zza}%
  \BibitemOpen
  \bibfield  {author} {\bibinfo {author} {\bibfnamefont {V.}~\bibnamefont
  {Chung}},\ }\href {\doibase 10.1103/PhysRev.140.B1110} {\bibfield  {journal}
  {\bibinfo  {journal} {Phys. Rev.}\ }\textbf {\bibinfo {volume} {140}},\
  \bibinfo {pages} {B1110} (\bibinfo {year} {1965})}\BibitemShut {NoStop}%
\bibitem [{\citenamefont {Kibble}(1968)}]{Kibble:1968a}%
  \BibitemOpen
  \bibfield  {author} {\bibinfo {author} {\bibfnamefont {T.}~\bibnamefont
  {Kibble}},\ }\href@noop {} {\bibfield  {journal} {\bibinfo  {journal} {J.
  Math. Phys.}\ }\textbf {\bibinfo {volume} {9}},\ \bibinfo {pages} {315}
  (\bibinfo {year} {1968})}\BibitemShut {NoStop}%
\bibitem [{\citenamefont {Kulish}\ and\ \citenamefont
  {Faddeev}(1970)}]{Kulish:1970ut}%
  \BibitemOpen
  \bibfield  {author} {\bibinfo {author} {\bibfnamefont {P.~P.}\ \bibnamefont
  {Kulish}}\ and\ \bibinfo {author} {\bibfnamefont {L.~D.}\ \bibnamefont
  {Faddeev}},\ }\href {\doibase 10.1007/BF01066485} {\bibfield  {journal}
  {\bibinfo  {journal} {Theor. Math. Phys.}\ }\textbf {\bibinfo {volume} {4}},\
  \bibinfo {pages} {745} (\bibinfo {year} {1970})},\ \bibinfo {note} {[Teor.
  Mat. Fiz. 4, 153(1970)]}\BibitemShut {NoStop}%
\bibitem [{\citenamefont {Horan}\ \emph {et~al.}(1998)\citenamefont {Horan},
  \citenamefont {Lavelle},\ and\ \citenamefont {McMullan}}]{Horan:1998im}%
  \BibitemOpen
  \bibfield  {author} {\bibinfo {author} {\bibfnamefont {R.}~\bibnamefont
  {Horan}}, \bibinfo {author} {\bibfnamefont {M.}~\bibnamefont {Lavelle}}, \
  and\ \bibinfo {author} {\bibfnamefont {D.}~\bibnamefont {McMullan}},\ }\href
  {\doibase 10.1007/BF02828927} {\bibfield  {journal} {\bibinfo  {journal}
  {Pramana}\ }\textbf {\bibinfo {volume} {51}},\ \bibinfo {pages} {317}
  (\bibinfo {year} {1998})},\ \Eprint {http://arxiv.org/abs/hep-th/9810089}
  {arXiv:hep-th/9810089 [hep-th]} \BibitemShut {NoStop}%
\bibitem [{\citenamefont {Bender}\ and\ \citenamefont
  {Klevansky}(2010)}]{Bender:2010hf}%
  \BibitemOpen
  \bibfield  {author} {\bibinfo {author} {\bibfnamefont {C.~M.}\ \bibnamefont
  {Bender}}\ and\ \bibinfo {author} {\bibfnamefont {S.~P.}\ \bibnamefont
  {Klevansky}},\ }\href {\doibase 10.1103/PhysRevLett.105.031601} {\bibfield
  {journal} {\bibinfo  {journal} {Phys. Rev. Lett.}\ }\textbf {\bibinfo
  {volume} {105}},\ \bibinfo {pages} {031601} (\bibinfo {year} {2010})},\
  \Eprint {http://arxiv.org/abs/1002.3253} {arXiv:1002.3253 [hep-th]}
  \BibitemShut {NoStop}%
\bibitem [{\citenamefont {Lippmann}\ and\ \citenamefont
  {Schwinger}(1950)}]{Lippmann:1950zz}%
  \BibitemOpen
  \bibfield  {author} {\bibinfo {author} {\bibfnamefont {B.~A.}\ \bibnamefont
  {Lippmann}}\ and\ \bibinfo {author} {\bibfnamefont {J.}~\bibnamefont
  {Schwinger}},\ }\href {\doibase 10.1103/PhysRev.79.469} {\bibfield  {journal}
  {\bibinfo  {journal} {Phys. Rev.}\ }\textbf {\bibinfo {volume} {79}},\
  \bibinfo {pages} {469} (\bibinfo {year} {1950})}\BibitemShut {NoStop}%
\bibitem [{\citenamefont {Weinberg}(2005)}]{Weinberg:1995mt}%
  \BibitemOpen
  \bibfield  {author} {\bibinfo {author} {\bibfnamefont {S.}~\bibnamefont
  {Weinberg}},\ }\href@noop {} {\emph {\bibinfo {title} {{The Quantum theory of
  fields. Vol. 1: Foundations}}}}\ (\bibinfo  {publisher} {Cambridge University
  Press},\ \bibinfo {year} {2005})\BibitemShut {NoStop}%
\bibitem [{\citenamefont {Bjorken}\ and\ \citenamefont
  {Drell}(1964)}]{Bjorken:1964}%
  \BibitemOpen
  \bibfield  {author} {\bibinfo {author} {\bibfnamefont {J.~D.}\ \bibnamefont
  {Bjorken}}\ and\ \bibinfo {author} {\bibfnamefont {S.~D.}\ \bibnamefont
  {Drell}},\ }\href@noop {} {\emph {\bibinfo {title} {{Relativistic quantum
  mechanics}}}}\ (\bibinfo  {publisher} {McGraw-Hill, New York},\ \bibinfo
  {year} {1964})\BibitemShut {NoStop}%
\bibitem [{\citenamefont {Lavelle}\ \emph {et~al.}(2013)\citenamefont
  {Lavelle}, \citenamefont {McMullan},\ and\ \citenamefont
  {Raddadi}}]{Lavelle:2013wx}%
  \BibitemOpen
  \bibfield  {author} {\bibinfo {author} {\bibfnamefont {M.}~\bibnamefont
  {Lavelle}}, \bibinfo {author} {\bibfnamefont {D.}~\bibnamefont {McMullan}}, \
  and\ \bibinfo {author} {\bibfnamefont {M.}~\bibnamefont {Raddadi}},\ }\href
  {\doibase 10.1103/PhysRevD.87.085024} {\bibfield  {journal} {\bibinfo
  {journal} {Phys. Rev.}\ }\textbf {\bibinfo {volume} {D87}},\ \bibinfo {pages}
  {085024} (\bibinfo {year} {2013})},\ \Eprint {http://arxiv.org/abs/1301.3072}
  {arXiv:1301.3072 [hep-ph]} \BibitemShut {NoStop}%
\bibitem [{\citenamefont {Nikishov}\ and\ \citenamefont
  {Ritus}(1964{\natexlab{a}})}]{Nikishov:1964zza}%
  \BibitemOpen
  \bibfield  {author} {\bibinfo {author} {\bibfnamefont {A.~I.}\ \bibnamefont
  {Nikishov}}\ and\ \bibinfo {author} {\bibfnamefont {V.~I.}\ \bibnamefont
  {Ritus}},\ }\href@noop {} {\bibfield  {journal} {\bibinfo  {journal} {Sov.
  Phys. JETP}\ }\textbf {\bibinfo {volume} {19}},\ \bibinfo {pages} {529}
  (\bibinfo {year} {1964}{\natexlab{a}})},\ \bibinfo {note} {[Zh. Eksp. Teor.
  Fiz. \textbf{46}, 776 (1963)]}\BibitemShut {NoStop}%
\bibitem [{\citenamefont {Nikishov}\ and\ \citenamefont
  {Ritus}(1964{\natexlab{b}})}]{Nikishov:1964zz}%
  \BibitemOpen
  \bibfield  {author} {\bibinfo {author} {\bibfnamefont {A.~I.}\ \bibnamefont
  {Nikishov}}\ and\ \bibinfo {author} {\bibfnamefont {V.~I.}\ \bibnamefont
  {Ritus}},\ }\href@noop {} {\bibfield  {journal} {\bibinfo  {journal} {Sov.
  Phys. JETP}\ }\textbf {\bibinfo {volume} {19}},\ \bibinfo {pages} {1191}
  (\bibinfo {year} {1964}{\natexlab{b}})},\ \bibinfo {note} {[Zh. Eksp. Teor.
  Fiz. \textbf{46}, 1768 (1964)]}\BibitemShut {NoStop}%
\bibitem [{\citenamefont {Narozhnyi}\ \emph {et~al.}(1965)\citenamefont
  {Narozhnyi}, \citenamefont {Nikishov},\ and\ \citenamefont
  {Ritus}}]{Narozhnyi:1965}%
  \BibitemOpen
  \bibfield  {author} {\bibinfo {author} {\bibfnamefont {N.~B.}\ \bibnamefont
  {Narozhnyi}}, \bibinfo {author} {\bibfnamefont {A.~I.}\ \bibnamefont
  {Nikishov}}, \ and\ \bibinfo {author} {\bibfnamefont {V.~I.}\ \bibnamefont
  {Ritus}},\ }\href@noop {} {\bibfield  {journal} {\bibinfo  {journal} {Sov.
  Phys. JETP}\ }\textbf {\bibinfo {volume} {20}},\ \bibinfo {pages} {622}
  (\bibinfo {year} {1965})},\ \bibinfo {note} {[Zh.\ Eksp.\ Teor.\ Fiz.
  \textbf{47}, 930 (1964)]}\BibitemShut {NoStop}%
\bibitem [{\citenamefont {Nikishov}\ and\ \citenamefont
  {Ritus}(1965)}]{Nikishov:1965}%
  \BibitemOpen
  \bibfield  {author} {\bibinfo {author} {\bibfnamefont {A.~I.}\ \bibnamefont
  {Nikishov}}\ and\ \bibinfo {author} {\bibfnamefont {V.~I.}\ \bibnamefont
  {Ritus}},\ }\href@noop {} {\bibfield  {journal} {\bibinfo  {journal} {Sov.
  Phys. JETP}\ }\textbf {\bibinfo {volume} {20}},\ \bibinfo {pages} {757}
  (\bibinfo {year} {1965})},\ \bibinfo {note} {[Zh. Eksp. Teor. Fiz.
  \textbf{46}, 1768 (1964)]}\BibitemShut {NoStop}%
\bibitem [{\citenamefont {Berestetskii}\ \emph {et~al.}(1982)\citenamefont
  {Berestetskii}, \citenamefont {Lifshitz},\ and\ \citenamefont
  {Pitaevskii}}]{landau4}%
  \BibitemOpen
  \bibfield  {author} {\bibinfo {author} {\bibfnamefont {V.~B.}\ \bibnamefont
  {Berestetskii}}, \bibinfo {author} {\bibfnamefont {E.~M.}\ \bibnamefont
  {Lifshitz}}, \ and\ \bibinfo {author} {\bibfnamefont {L.~P.}\ \bibnamefont
  {Pitaevskii}},\ }\href@noop {} {\emph {\bibinfo {title} {Quantum
  Electrodynamics}}}\ (\bibinfo  {publisher} {Butterworth-Heinemann},\ \bibinfo
  {address} {Oxford},\ \bibinfo {year} {1982})\BibitemShut {NoStop}%
\bibitem [{\citenamefont {Harvey}\ \emph {et~al.}(2009)\citenamefont {Harvey},
  \citenamefont {Heinzl},\ and\ \citenamefont {Ilderton}}]{Harvey:2009ry}%
  \BibitemOpen
  \bibfield  {author} {\bibinfo {author} {\bibfnamefont {C.}~\bibnamefont
  {Harvey}}, \bibinfo {author} {\bibfnamefont {T.}~\bibnamefont {Heinzl}}, \
  and\ \bibinfo {author} {\bibfnamefont {A.}~\bibnamefont {Ilderton}},\ }\href
  {\doibase 10.1103/PhysRevA.79.063407} {\bibfield  {journal} {\bibinfo
  {journal} {Phys. Rev.}\ }\textbf {\bibinfo {volume} {A79}},\ \bibinfo {pages}
  {063407} (\bibinfo {year} {2009})},\ \Eprint {http://arxiv.org/abs/0903.4151}
  {arXiv:0903.4151 [hep-ph]} \BibitemShut {NoStop}%
\bibitem [{\citenamefont {Seipt}\ \emph {et~al.}(2017)\citenamefont {Seipt},
  \citenamefont {Heinzl}, \citenamefont {Marklund},\ and\ \citenamefont
  {Bulanov}}]{Seipt:2016fyu}%
  \BibitemOpen
  \bibfield  {author} {\bibinfo {author} {\bibfnamefont {D.}~\bibnamefont
  {Seipt}}, \bibinfo {author} {\bibfnamefont {T.}~\bibnamefont {Heinzl}},
  \bibinfo {author} {\bibfnamefont {M.}~\bibnamefont {Marklund}}, \ and\
  \bibinfo {author} {\bibfnamefont {S.~S.}\ \bibnamefont {Bulanov}},\ }\href
  {\doibase 10.1103/PhysRevLett.118.154803} {\bibfield  {journal} {\bibinfo
  {journal} {Phys. Rev. Lett.}\ }\textbf {\bibinfo {volume} {118}},\ \bibinfo
  {pages} {154803} (\bibinfo {year} {2017})},\ \Eprint
  {http://arxiv.org/abs/1605.00633} {arXiv:1605.00633 [hep-ph]} \BibitemShut
  {NoStop}%
\bibitem [{\citenamefont {Neville}\ and\ \citenamefont
  {Rohrlich}(1971)}]{Neville:1971uc}%
  \BibitemOpen
  \bibfield  {author} {\bibinfo {author} {\bibfnamefont {R.~A.}\ \bibnamefont
  {Neville}}\ and\ \bibinfo {author} {\bibfnamefont {F.}~\bibnamefont
  {Rohrlich}},\ }\href {\doibase 10.1103/PhysRevD.3.1692} {\bibfield  {journal}
  {\bibinfo  {journal} {Phys. Rev.}\ }\textbf {\bibinfo {volume} {D3}},\
  \bibinfo {pages} {1692} (\bibinfo {year} {1971})}\BibitemShut {NoStop}%
\bibitem [{\citenamefont {Ji}\ and\ \citenamefont {Rey}(1996)}]{Ji:1995ft}%
  \BibitemOpen
  \bibfield  {author} {\bibinfo {author} {\bibfnamefont {C.-R.}\ \bibnamefont
  {Ji}}\ and\ \bibinfo {author} {\bibfnamefont {S.-J.}\ \bibnamefont {Rey}},\
  }\href {\doibase 10.1103/PhysRevD.53.5815} {\bibfield  {journal} {\bibinfo
  {journal} {Phys. Rev.}\ }\textbf {\bibinfo {volume} {D53}},\ \bibinfo {pages}
  {5815} (\bibinfo {year} {1996})},\ \Eprint
  {http://arxiv.org/abs/hep-ph/9505420} {arXiv:hep-ph/9505420 [hep-ph]}
  \BibitemShut {NoStop}%
\bibitem [{\citenamefont {Tomaras}\ \emph {et~al.}(2000)\citenamefont
  {Tomaras}, \citenamefont {Tsamis},\ and\ \citenamefont
  {Woodard}}]{Tomaras:2000ag}%
  \BibitemOpen
  \bibfield  {author} {\bibinfo {author} {\bibfnamefont {T.~N.}\ \bibnamefont
  {Tomaras}}, \bibinfo {author} {\bibfnamefont {N.~C.}\ \bibnamefont {Tsamis}},
  \ and\ \bibinfo {author} {\bibfnamefont {R.~P.}\ \bibnamefont {Woodard}},\
  }\href {\doibase 10.1103/PhysRevD.62.125005} {\bibfield  {journal} {\bibinfo
  {journal} {Phys. Rev.}\ }\textbf {\bibinfo {volume} {D62}},\ \bibinfo {pages}
  {125005} (\bibinfo {year} {2000})},\ \Eprint
  {http://arxiv.org/abs/hep-ph/0007166} {arXiv:hep-ph/0007166 [hep-ph]}
  \BibitemShut {NoStop}%
\bibitem [{\citenamefont {Tomaras}\ \emph {et~al.}(2001)\citenamefont
  {Tomaras}, \citenamefont {Tsamis},\ and\ \citenamefont
  {Woodard}}]{Tomaras:2001vs}%
  \BibitemOpen
  \bibfield  {author} {\bibinfo {author} {\bibfnamefont {T.~N.}\ \bibnamefont
  {Tomaras}}, \bibinfo {author} {\bibfnamefont {N.~C.}\ \bibnamefont {Tsamis}},
  \ and\ \bibinfo {author} {\bibfnamefont {R.~P.}\ \bibnamefont {Woodard}},\
  }\href {\doibase 10.1088/1126-6708/2001/11/008} {\bibfield  {journal}
  {\bibinfo  {journal} {JHEP}\ }\textbf {\bibinfo {volume} {11}},\ \bibinfo
  {pages} {008} (\bibinfo {year} {2001})},\ \Eprint
  {http://arxiv.org/abs/hep-th/0108090} {arXiv:hep-th/0108090 [hep-th]}
  \BibitemShut {NoStop}%
\bibitem [{\citenamefont {Ilderton}(2014)}]{Ilderton:2014mla}%
  \BibitemOpen
  \bibfield  {author} {\bibinfo {author} {\bibfnamefont {A.}~\bibnamefont
  {Ilderton}},\ }\href {\doibase 10.1007/JHEP09(2014)166} {\bibfield  {journal}
  {\bibinfo  {journal} {JHEP}\ }\textbf {\bibinfo {volume} {09}},\ \bibinfo
  {pages} {166} (\bibinfo {year} {2014})},\ \Eprint
  {http://arxiv.org/abs/1406.1513} {arXiv:1406.1513 [hep-th]} \BibitemShut
  {NoStop}%
\bibitem [{\citenamefont {Bakker}\ \emph {et~al.}(2014)\citenamefont {Bakker}
  \emph {et~al.}}]{Bakker:2013cea}%
  \BibitemOpen
  \bibfield  {author} {\bibinfo {author} {\bibfnamefont {B.~L.~G.}\
  \bibnamefont {Bakker}} \emph {et~al.},\ }\href {\doibase
  10.1016/j.nuclphysbps.2014.05.004} {\bibfield  {journal} {\bibinfo  {journal}
  {Nucl. Phys. Proc. Suppl.}\ }\textbf {\bibinfo {volume} {251-252}},\ \bibinfo
  {pages} {165} (\bibinfo {year} {2014})},\ \Eprint
  {http://arxiv.org/abs/1309.6333} {arXiv:1309.6333 [hep-ph]} \BibitemShut
  {NoStop}%
\bibitem [{\citenamefont {Wigner}(1939)}]{Wigner:1939cj}%
  \BibitemOpen
  \bibfield  {author} {\bibinfo {author} {\bibfnamefont {E.~P.}\ \bibnamefont
  {Wigner}},\ }\href {\doibase 10.2307/1968551} {\bibfield  {journal} {\bibinfo
   {journal} {Annals Math.}\ }\textbf {\bibinfo {volume} {40}},\ \bibinfo
  {pages} {149} (\bibinfo {year} {1939})},\ \bibinfo {note} {[Reprint: Nucl.
  Phys. Proc. Suppl. 6, 9 (1989)]}\BibitemShut {NoStop}%
\bibitem [{\citenamefont {King}\ and\ \citenamefont
  {Heinzl}(2016)}]{King:2015tba}%
  \BibitemOpen
  \bibfield  {author} {\bibinfo {author} {\bibfnamefont {B.}~\bibnamefont
  {King}}\ and\ \bibinfo {author} {\bibfnamefont {T.}~\bibnamefont {Heinzl}},\
  }\href {\doibase 10.1017/hpl.2016.1} {\bibfield  {journal} {\bibinfo
  {journal} {High Power Laser Science and Engineering}\ }\textbf {\bibinfo
  {volume} {4}},\ \bibinfo {pages} {e5} (\bibinfo {year} {2016})},\ \Eprint
  {http://arxiv.org/abs/1510.08456} {arXiv:1510.08456 [hep-ph]} \BibitemShut
  {NoStop}%
\bibitem [{\citenamefont {Skoromnik}\ \emph {et~al.}(2015)\citenamefont
  {Skoromnik}, \citenamefont {Feranchuk}, \citenamefont {Lu},\ and\
  \citenamefont {Keitel}}]{Skoromnik:2015hya}%
  \BibitemOpen
  \bibfield  {author} {\bibinfo {author} {\bibfnamefont {O.~D.}\ \bibnamefont
  {Skoromnik}}, \bibinfo {author} {\bibfnamefont {I.~D.}\ \bibnamefont
  {Feranchuk}}, \bibinfo {author} {\bibfnamefont {D.~V.}\ \bibnamefont {Lu}}, \
  and\ \bibinfo {author} {\bibfnamefont {C.~H.}\ \bibnamefont {Keitel}},\
  }\href {\doibase 10.1103/PhysRevD.92.125019} {\bibfield  {journal} {\bibinfo
  {journal} {Phys. Rev.}\ }\textbf {\bibinfo {volume} {D92}},\ \bibinfo {pages}
  {125019} (\bibinfo {year} {2015})},\ \Eprint
  {http://arxiv.org/abs/1506.07025} {arXiv:1506.07025 [quant-ph]} \BibitemShut
  {NoStop}%
\bibitem [{\citenamefont {Pauli}\ and\ \citenamefont
  {Brodsky}(1985{\natexlab{a}})}]{Pauli:1985pv}%
  \BibitemOpen
  \bibfield  {author} {\bibinfo {author} {\bibfnamefont {H.~C.}\ \bibnamefont
  {Pauli}}\ and\ \bibinfo {author} {\bibfnamefont {S.~J.}\ \bibnamefont
  {Brodsky}},\ }\href {\doibase 10.1103/PhysRevD.32.1993} {\bibfield  {journal}
  {\bibinfo  {journal} {Phys. Rev.}\ }\textbf {\bibinfo {volume} {D32}},\
  \bibinfo {pages} {1993} (\bibinfo {year} {1985}{\natexlab{a}})}\BibitemShut
  {NoStop}%
\bibitem [{\citenamefont {Pauli}\ and\ \citenamefont
  {Brodsky}(1985{\natexlab{b}})}]{Pauli:1985ps}%
  \BibitemOpen
  \bibfield  {author} {\bibinfo {author} {\bibfnamefont {H.~C.}\ \bibnamefont
  {Pauli}}\ and\ \bibinfo {author} {\bibfnamefont {S.~J.}\ \bibnamefont
  {Brodsky}},\ }\href {\doibase 10.1103/PhysRevD.32.2001} {\bibfield  {journal}
  {\bibinfo  {journal} {Phys. Rev.}\ }\textbf {\bibinfo {volume} {D32}},\
  \bibinfo {pages} {2001} (\bibinfo {year} {1985}{\natexlab{b}})}\BibitemShut
  {NoStop}%
\bibitem [{\citenamefont {Zhao}\ \emph {et~al.}(2013)\citenamefont {Zhao},
  \citenamefont {Ilderton}, \citenamefont {Maris},\ and\ \citenamefont
  {Vary}}]{Zhao:2013cma}%
  \BibitemOpen
  \bibfield  {author} {\bibinfo {author} {\bibfnamefont {X.}~\bibnamefont
  {Zhao}}, \bibinfo {author} {\bibfnamefont {A.}~\bibnamefont {Ilderton}},
  \bibinfo {author} {\bibfnamefont {P.}~\bibnamefont {Maris}}, \ and\ \bibinfo
  {author} {\bibfnamefont {J.~P.}\ \bibnamefont {Vary}},\ }\href {\doibase
  10.1103/PhysRevD.88.065014} {\bibfield  {journal} {\bibinfo  {journal} {Phys.
  Rev.}\ }\textbf {\bibinfo {volume} {D88}},\ \bibinfo {pages} {065014}
  (\bibinfo {year} {2013})},\ \Eprint {http://arxiv.org/abs/1303.3273}
  {arXiv:1303.3273 [nucl-th]} \BibitemShut {NoStop}%
\bibitem [{\citenamefont {Bergou}\ and\ \citenamefont
  {Varro}(1980)}]{Bergou:1980cm}%
  \BibitemOpen
  \bibfield  {author} {\bibinfo {author} {\bibfnamefont {J.}~\bibnamefont
  {Bergou}}\ and\ \bibinfo {author} {\bibfnamefont {S.}~\bibnamefont {Varro}},\
  }\href@noop {} {\bibfield  {journal} {\bibinfo  {journal} {J. Phys. A}\
  }\textbf {\bibinfo {volume} {14}},\ \bibinfo {pages} {1469} (\bibinfo {year}
  {1980})}\BibitemShut {NoStop}%
\bibitem [{\citenamefont {Froehlich}(1952)}]{Froehlich:1952}%
  \BibitemOpen
  \bibfield  {author} {\bibinfo {author} {\bibfnamefont {H.}~\bibnamefont
  {Froehlich}},\ }\href@noop {} {\bibfield  {journal} {\bibinfo  {journal}
  {Proc. Roy. Soc. A}\ }\textbf {\bibinfo {volume} {215}},\ \bibinfo {pages}
  {291} (\bibinfo {year} {1952})}\BibitemShut {NoStop}%
\bibitem [{\citenamefont {Nunnenkamp}\ \emph {et~al.}(2011)\citenamefont
  {Nunnenkamp}, \citenamefont {B\o{}rkje},\ and\ \citenamefont
  {Girvin}}]{Nunnenkamp}%
  \BibitemOpen
  \bibfield  {author} {\bibinfo {author} {\bibfnamefont {A.}~\bibnamefont
  {Nunnenkamp}}, \bibinfo {author} {\bibfnamefont {K.}~\bibnamefont
  {B\o{}rkje}}, \ and\ \bibinfo {author} {\bibfnamefont {S.~M.}\ \bibnamefont
  {Girvin}},\ }\href {\doibase 10.1103/PhysRevLett.107.063602} {\bibfield
  {journal} {\bibinfo  {journal} {Phys. Rev. Lett.}\ }\textbf {\bibinfo
  {volume} {107}},\ \bibinfo {pages} {063602} (\bibinfo {year}
  {2011})}\BibitemShut {NoStop}%
\bibitem [{\citenamefont {Law}(1994)}]{Law-Hamiltonian}%
  \BibitemOpen
  \bibfield  {author} {\bibinfo {author} {\bibfnamefont {C.~K.}\ \bibnamefont
  {Law}},\ }\href {\doibase 10.1103/PhysRevA.49.433} {\bibfield  {journal}
  {\bibinfo  {journal} {Phys. Rev. A}\ }\textbf {\bibinfo {volume} {49}},\
  \bibinfo {pages} {433} (\bibinfo {year} {1994})}\BibitemShut {NoStop}%
\bibitem [{\citenamefont {Mati}(2017)}]{Mati:2016xdj}%
  \BibitemOpen
  \bibfield  {author} {\bibinfo {author} {\bibfnamefont {P.}~\bibnamefont
  {Mati}},\ }\href {\doibase 10.1103/PhysRevA.95.053852} {\bibfield  {journal}
  {\bibinfo  {journal} {Phys. Rev.}\ }\textbf {\bibinfo {volume} {A95}},\
  \bibinfo {pages} {053852} (\bibinfo {year} {2017})},\ \Eprint
  {http://arxiv.org/abs/1611.02255} {arXiv:1611.02255 [quant-ph]} \BibitemShut
  {NoStop}%
\bibitem [{\citenamefont {Armata}\ \emph {et~al.}(2017)\citenamefont {Armata},
  \citenamefont {Kim}, \citenamefont {Butera}, \citenamefont {Rizzuto},\ and\
  \citenamefont {Passante}}]{Armata:2017gmp}%
  \BibitemOpen
  \bibfield  {author} {\bibinfo {author} {\bibfnamefont {F.}~\bibnamefont
  {Armata}}, \bibinfo {author} {\bibfnamefont {M.~S.}\ \bibnamefont {Kim}},
  \bibinfo {author} {\bibfnamefont {S.}~\bibnamefont {Butera}}, \bibinfo
  {author} {\bibfnamefont {L.}~\bibnamefont {Rizzuto}}, \ and\ \bibinfo
  {author} {\bibfnamefont {R.}~\bibnamefont {Passante}},\ }\href {\doibase
  10.1103/PhysRevD.96.045007} {\bibfield  {journal} {\bibinfo  {journal} {Phys.
  Rev.}\ }\textbf {\bibinfo {volume} {D96}},\ \bibinfo {pages} {045007}
  (\bibinfo {year} {2017})},\ \Eprint {http://arxiv.org/abs/1707.01163}
  {arXiv:1707.01163 [quant-ph]} \BibitemShut {NoStop}%
\end{thebibliography}%

\end{document}